\documentclass[%
reprint,
superscriptaddress,
amsmath,amssymb,
aps,
pra,
]{revtex4-2}

\usepackage{graphicx}
\usepackage{dcolumn}
\usepackage{bm}
\usepackage{float}
\usepackage{hyperref}
\usepackage{braket}
\usepackage{physics}
\usepackage{xcolor}
\usepackage[normalem]{ulem}
\usepackage{upgreek}
\usepackage{soul}
\usepackage{amssymb}
\usepackage{amsmath}
\usepackage{multirow}
\usepackage{enumitem,kantlipsum}
\usepackage[caption=false]{subfig}
\usepackage{lineno,hyperref}

\begin{document}
	
	\preprint{APS/123-QED}
	
	\title{Diffraction of an Off-axis Vector-Beam by a Tilted Aperture}
	
	\author{Ghanasyam Remesh}
	\affiliation{Department of Condensed Matter Physics and Material Science, Tata Institute of Fundamental Research, Mumbai 400005, India}
	
	\author{Athira B S}
	\email{abs16rs013@iiserkol.ac.in}
	\affiliation{Center of Excellence in Space Sciences India, Indian Institute of Science Education and Research (IISER) Kolkata. Mohanpur 741246, India.}
	
	\author{Shyamal Gucchait}
	\affiliation{Department of Physical Sciences, Indian Institute of Science Education and Research (IISER) Kolkata. Mohanpur 741246, India.}
	
	\author{Ayan Banerjee}
	\affiliation{Department of Physical Sciences, Indian Institute of Science Education and Research (IISER) Kolkata. Mohanpur 741246, India.}
	\affiliation{Center of Excellence in Space Sciences India, Indian Institute of Science Education and Research (IISER) Kolkata. Mohanpur 741246, India.}
	
	\author{Nirmalya Ghosh}
	\affiliation{Department of Physical Sciences, Indian Institute of Science Education and Research (IISER) Kolkata. Mohanpur 741246, India.}
	\affiliation{Center of Excellence in Space Sciences India, Indian Institute of Science Education and Research (IISER) Kolkata. Mohanpur 741246, India.}
	
	\author{Subhasish Dutta Gupta}
	\email{sdghyderabad@gmail.com}
	\affiliation{Department of Physical Sciences, Indian Institute of Science Education and Research (IISER) Kolkata. Mohanpur 741246, India.}
	\affiliation{Tata Centre for Interdisciplinary Sciences, TIFRH, Hyderabad 500107, India}
	\affiliation{School of Physics, University of Hyderabad, Hyderabad 500046, India}

	\date{\today}
	
	\begin{abstract}
		Manifestations of  {orbital angular momentum induced effects} in the diffraction of a radially polarized vector beam by an off-axis tilted aperture are studied both experimentally and theoretically. Experiments were carried out to extract the degree of circular polarization, which was shown to be proportional to the on-axis component of the spin angular momentum density. We report a clear separation of the regions of dominance of the right and left circular polarizations {associated with positive and negative topological charges respectively}, which is reminiscent of the standard {vortex-induced transverse} shift, albeit in the diffraction scenario. The experimental results are supported by model simulations and the agreement is quite satisfactory. The results are useful to appreciate the {orbit}-orbit related effects due to unavoidable misalignment problems (especially for vortex beams).
	\end{abstract}
	
	\maketitle
	
	\section{\label{sec:level1}Introduction}
	
It is now well understood that the angular momentum of light can have contributions from two factors. One depends on the polarization of the field, and is related to the spin angular momentum (SAM). The other, drawing its origin from the spatial distribution of the field, being independent of the beam polarization, is related to orbital angular momentum (OAM). SAM is an intrinsic property of the field, i.e, independent of the choice of axis about which the angular momentum is calculated \cite{Padgett:2002}. For a circularly polarized light, the SAM per photon takes the value of $\pm \hbar$. Similarly, Allen \textit{et. al.} \cite{Allen:1992} showed that for Laguerre Gaussian (LG) beams, each photon has an OAM of $l\hbar$, where $l$ is the topological charge of the vortex. However, in contrast to SAM, the OAM can either be intrinsic or extrinsic in nature, depending on the geometry of the system. The OAM of an LG beam causes a particle to rotate around the beam axis, and a SAM would cause a birefringent material to spin about its own axis, and hence can be used as an `optical spanner' \cite{Padgett:2016a}. Thus, it should be noted that the separation of angular momentum into OAM and SAM is not just a theoretical result.  
\par
Several interactions happen between this intrinsic and extrinsic angular momentum, thereby giving rise to  spin-orbit {and orbit-orbit interaction (SOI and OOI, respectively).} The interactions can be divided into three categories: between SAM and extrinsic OAM, SAM and intrinsic OAM and  intrinsic and extrinsic OAM {(related to OOI)} \cite{Bliokh:2015,rodriguez2010optical,bliokh2008geometrodynamics}. This paper deals with the {last, where we have a separation in the spatial profile depending on the sign of the topological charge}. Owing to both fundamental interests and its potential applications the studies on {OOI} of light and its manifestations are currently attracting a lot of attention. \cite{Bliokh:2015,bliokh2011spin,zhao2007spin,marrucci2006optical,schwartz2006conservation}.
\par
 {Several authors  study the effects of Goos–Hänchen and Imbert–Fedorov shift in Gaussian and vortex beams \cite{Bliokh2009,Fedoseyev2008,Bliokh:2006,aiello2009transverse,Dasgupta2006,Loeffler2012}. A detailed description of these phenomena is covered in \cite{Bliokh:2013}.} There has been extensive research on intrinsic and extrinsic properties of the angular momentum in the context of LG beams \cite{Padgett:2002,Padgett:2000}. {It is well understood that an LG beam falling on an aperture  whose center is shifted from the beam center  (without any tilt) leads to a change in the distribution of the SAM density inside the aperture. A recent paper by Taira and Zhang\cite{Taira:2017} has a focus on shifted aperture-beam system without any tilt in order to bring out the ability of such systems to measure the vortex charge of (higher order) LG beams by mixing it with a reference beam.} However, there exists no such studies on the general vector beams falling on a tilted aperture. A radial vector beams can be considered to be a superposition of right and left circularly polarized LG beams of equal intensities and opposite OAM densities. {Vortex charge dependence of the diffracted pattern projected onto the circular basis manifests itself as OAM induced transverse shift of the LCP and RCP components, which is clearly a manifestation of OOI} \cite{bliokh2008geometrodynamics,hosten2008observation,aiello2009transverse,zhou2012identifying,AthiraB2021}.  
\par
In this paper, we study the effects {of both tilt and shift on orbit-orbit coupling due to diffraction of a vector beam by an off-centered tilted aperture. In particular, we explore the tilt- and shift- dependence of the topological charge-induced transverse shifts of the circular polarization components.} In the experiment, we perform Stokes imaging to extract the degree of circular polarization for a radially polarized vector beam diffracted by a tilted circular aperture. Further, we do numerical calculations of the same. For a comparison of our experimental results with the theoretically computed SAM, we establish a link between the experimentally measured Stokes parameter and the spin-angular momentum density. We also discuss in detail the procedure with due attention to the complications arising from  the tilt and shift of the  aperture. Since it is almost impossible to center a vortex beam exactly around a point, we assume in our numerical calculations that the beam is off-centered by a small distance. The high degree of resemblance between the experimentally observed and simulated results suggest that we can indeed trace the origins of the observed behavior to the tilt of the aperture and the shift of the beam. It should be noted that the finite aperture thickness has not been taken into account in any of the numerical calculations.
	\section{\label{sec:level2}Formulation of the problem}	
	\begin{figure}[t]
		\includegraphics[width=0.95\columnwidth]{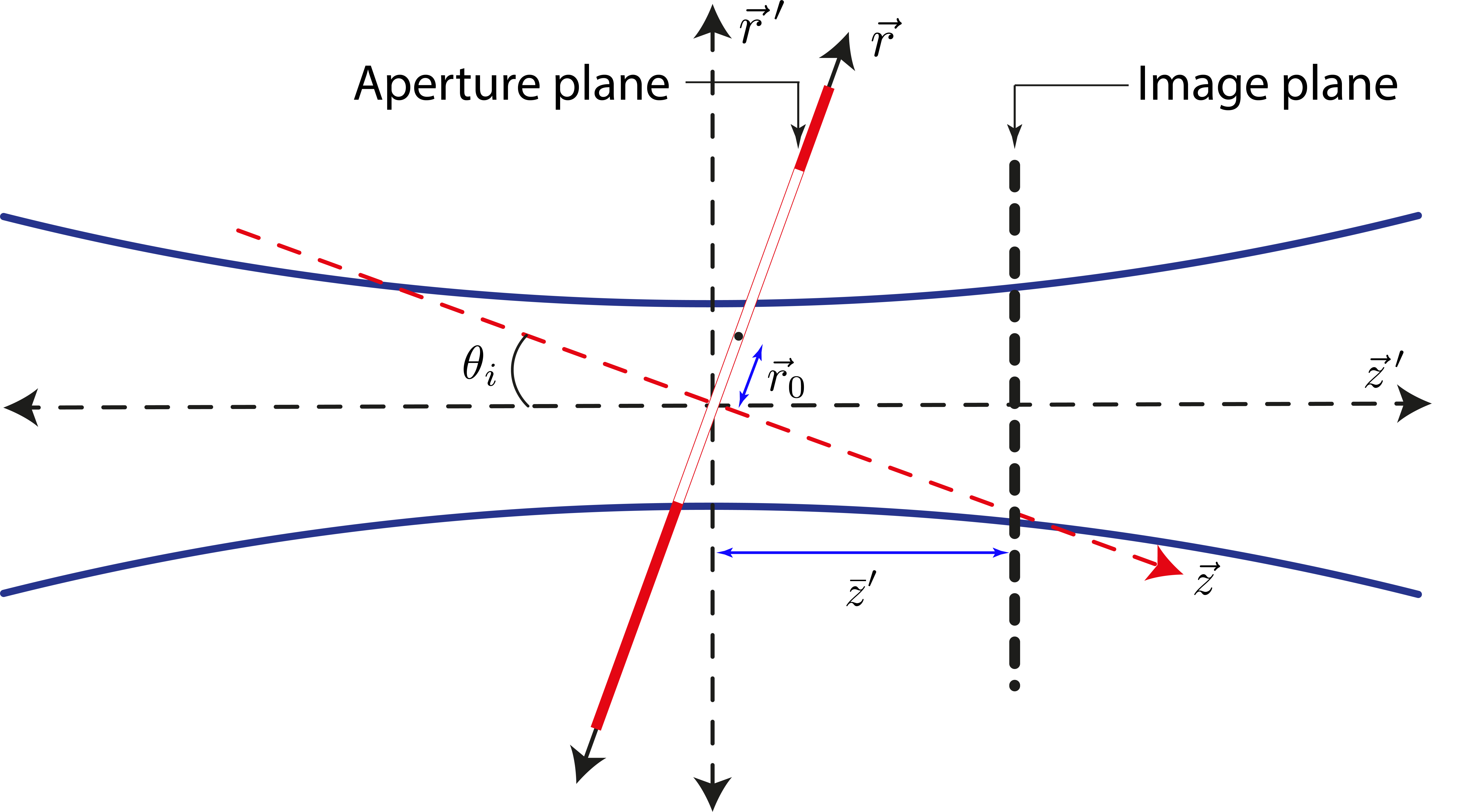}
		\caption{Schematics of the setup used for the numerical calculations. The orientation of the different axes as defined in the text are shown. The primed coordinate system is attached to the beam, where as the un-primed coordinate system describes the aperture plane. The rotation between the coordinate systems is assumed to be in the $x-z$ plane. Hence, both $\hat x$ and $\hat x'$ would lie on the plane of the this paper, while $\hat y=\hat y'$ projects out of the this plane. The center of the aperture is shifted from the origin by $\boldsymbol{r_0}$ as shown in the figure.}
		\label{Fig:Model}
	\end{figure}
	For a general EM field, the linear and angular momentum of light can be broken up into two parts, spin and orbital angular momentum, as mentioned
	in the introduction.  Recall that the angular momentum density of light is given by $\boldsymbol{r}\times \boldsymbol{p}$, where $\boldsymbol{p}$ is the linear momentum density of light given by $ \boldsymbol{p} = \frac{\epsilon_0}{2}\mbox{Re } [\boldsymbol{E^*}\times\boldsymbol{B}]$.  Using Maxwell's equations and vector identities, we can break up $\boldsymbol{p}$ into $\boldsymbol{p_{orb}}$ and $\boldsymbol{p_{sp}}$. The structures for the spin and orbital parts of the linear momentum densities are,
	\begin{eqnarray}
		\boldsymbol{p} &=& \boldsymbol{p_{orb}} + \boldsymbol{p_{sp}},\\
		\mbox{where }\boldsymbol{p}_{orb}	&=&  \frac{\epsilon_0}{2\omega}\mbox{Im } [\boldsymbol E^*\,.\,\boldsymbol \nabla \boldsymbol E],\label{eq:p_orb}\\
		\boldsymbol{p}_{sp}	&=&  \frac{\epsilon_0}{2\omega}\mbox{Im } [ \frac{1}{2}\boldsymbol \nabla\times(\boldsymbol E^*\times\boldsymbol E)].
	\end{eqnarray}
	\noindent In eq. \ref{eq:p_orb}, we have used the notation, $[\boldsymbol A\,.\,\boldsymbol\nabla \boldsymbol B]_j $= $A_i\left( \frac{\partial B_i}{\partial x_j} \right)$. Accordingly, the angular momentum density can also be broken up into two parts.
	\begin{eqnarray}
		\boldsymbol j &=& \boldsymbol j_{OAM} + \boldsymbol j_{SAM} = \boldsymbol r\times \boldsymbol{p}_{orb} + \boldsymbol r\times \boldsymbol{p}_{sp}.
	\end{eqnarray}
	\noindent Here, the first term $j_{OAM}$  leads to the intrinsic or extrinsic spin depending on the geometry and is independent of the polarization of the field. The second term represents the spin angular momentum density of light $\boldsymbol{j}_{SAM}$, which is always intrinsic in nature. Our focus is mainly on SAM and we show that it can be related to one of the Stokes parameters at the end of this section.  Assuming the electromagnetic field to be transverse, we can rewrite the expression for spin angular momentum density by integrating over space to find the total spin angular momentum using the boundary condition that the fields go to zero at the edge of the aperture and beyond. Finally, we have \cite{Allen:1994}
	\begin{eqnarray}
		\boldsymbol{j}_{SAM}  = \frac{\epsilon_0}{2\omega}\mbox{Im }[\boldsymbol E^* \times\boldsymbol E] = -\frac{\epsilon_0}{\omega} \mbox{Im } [E_y^*E_x].
	\end{eqnarray}
	\noindent For a field that satisfies the paraxial equation, such as for the LG beam, it can be seen that the ratio of SAM density to photon density depends on the polarization of the field, and the corresponding value for $\boldsymbol{j}_{OAM}$ depends on $l$. Hence, the physical origins of the two spins are clearly different.
	\par
	In what follows, we show that $j_{SAM}$ can directly be extracted from the final Stokes parameter, $V$. It is well known that for an an electric field vector given by $\boldsymbol{E} = E_x \boldsymbol{\hat x} + E_y \boldsymbol{\hat y}$, the Stokes parameter, $I$ and $V$ are given by,
	\begin{eqnarray}
		I &=& |E_x|^2 + |E_y|^2,\\
		V &=& 2 \mbox{Im } [E_y^*E_x].
	\end{eqnarray}
	\noindent Thus, both SAM and $V$ have the same structure, and thus, by means of  Stokes mapping we can reveal the character of spin angular momentum of the diffracted beam. The numerical computation results can then be compared with the measured $V$.
	\par
	In what follows, we give a brief sketch of our theoretical calculations and describe the corresponding experimental setup.  Our experiment directly measures the values of $V$ and $I$, as will be detailed in section \ref{Sec:Experimental_realization}.
	\subsection{Theoretical approach}
	In contrast to the existing studies \cite{Padgett:2002,Taira:2017}, this paper incorporates the tilt of the aperture with respect to the beam axis.  We study the diffraction of a radial vector beam propagating in the $\boldsymbol{\hat z'}$ direction and falling on a tilted aperture, as shown in Fig. \ref{Fig:Model}. The image is captured in a plane that is taken to be perpendicular to the direction of propagation of the beam. For ease of calculations, we assume the plane of the aperture to pass through the center of the beam waist. As mentioned before, the center of the aperture is assumed to be shifted by $\boldsymbol{r_0}=x_0\, \boldsymbol{\hat x}+y_0\,\boldsymbol{ \hat y}$ from (0,0), i.e, the center of the beam waist. Here $\boldsymbol{\hat x}$ and $\boldsymbol{\hat y}$ refers to unit vectors that lie on the aperture plane, and $\boldsymbol{\hat z}$ is the normal to this plane, i.e, the unprimed coordinate system is attached to the aperture. Similarly, $\boldsymbol{\hat x'}$ and $\boldsymbol{\hat y'}$ lie on the image plane, and $\boldsymbol{\hat z'}$ is the normal to this plane, i.e, the primed coordinate system is assigned to the beam. We assume the origins of both coordinate systems to be coincident. The angle between $\boldsymbol{\hat z}$ and $\boldsymbol{\hat z'}$ is taken to be $\theta_i$.  
	\par
	Vector beams have been studied extensively and the analytical expression for the electric field can be written as \cite{AM:Book},
	\begin{eqnarray}
		\boldsymbol{E}(x',y',z') &=& E(r',z') \, [\cos(\theta+\phi)\,\boldsymbol{\hat x'} \notag\\ 
		&& + \sin(\theta+\phi)\,\exp(i\delta) \,\boldsymbol{\hat y'}], \label{eq_VB}
	\end{eqnarray}
	\noindent where,
	\begin{eqnarray}
		E(r',z') &=& \frac{2\sqrt{2}\,\omega_0}{\omega^2(z')} \,r'\, \exp\left(-\frac{{r'^2}}{\omega^2(z')}\right)\,
		\exp\left(i\frac{kr'^2}{2R(z')}\right)\notag\notag \\
		&&\exp{\left(-i\psi(z')\right)} \, \exp{(ikz')},
	\end{eqnarray}
	and $r'^2 = x'^2+y'^2$ and $\phi = \tan^{-1} \frac{y'}{x'}$. We introduce nominal deviation from the radial vector beam to account for the imperfections in the experimental scenario. Thus, $\theta$ and $\delta$  in eq. \ref{eq_VB} account for ellipticity of the beam. Note that for $\theta=\delta=0$, the beam polarization is in the radial direction.
	\par
	Let a vector beam as described by eq. \ref{eq_VB} fall on the tilted aperture as shown in Fig. \ref{Fig:Model}. Our goal is to calculate the diffraction pattern formed by this aperture at a screen located at a distance of $z'$. The method of using angular spectrum decomposition to propagate the field at an angle to the aperture plane has been discussed in detail in \cite{Matsushima:2003}. In what follows, we lay out the procedure and the mathematics involved in calculating the diffraction pattern. We also briefly discuss the complications that the titled aperture problem presents, as opposed to the case without the tilt.	
	\par
	We start with the field distribution in the opening of the aperture using eq. \ref{eq_VB} and the transformation equations between the two coordinate systems. For any point ($x$, $y$, $z=0$) which lies on the aperture plane, a simple coordinate rotation from the unprimed to the primed systems yields the relations,
	\begin{eqnarray}
		x' = x\cos\theta_i,\quad  y  = y',\quad z' = x\sin\theta_i \label{eq_x_transform}.
	\end{eqnarray}
	The  field at any point inside the aperture can be evaluated by using eq. \ref{eq_x_transform} and eq. \ref{eq_VB},
	\begin{eqnarray}
		\boldsymbol{E}(x,y,z=0) &=&  E(r',x\sin\theta_i)\,  [\cos(\theta+\phi)\,\boldsymbol{\hat x'} \notag\\
		&&+ \sin(\theta+\phi)\,\exp(i\delta) \,\boldsymbol{\hat y'}] \label{eq_VB_trans}, 
	\end{eqnarray}
	where $r'^2 = ( x\cos\theta_i)^2+y^2$ and $ \phi = \tan^{-1} \frac{y}{ x\cos\theta_i}$.
	The aperture in numerically implemented such that every point on the aperture plane that lies inside a circle of radius $a$  defined by $(x-x_0)^2+(y-y_0)^2 < a^2$  retains its field value as given by eq. \ref{eq_VB_trans}, whereas any point on or outside the boundary of the circle is assumed to have a zero field. It should be noted that this is exactly the same boundary condition as in Fresnel–Kirchhoff diffraction theory.
	\begin{figure}[t]
		\includegraphics[width=1 \columnwidth]{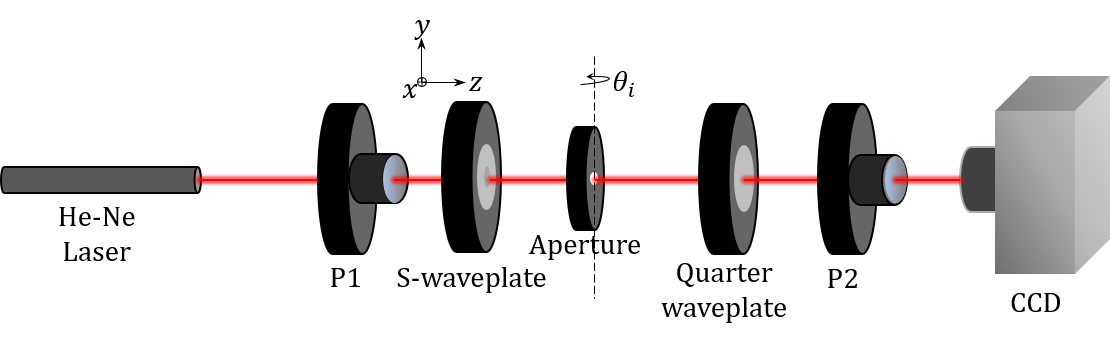}
		\caption{A schematic of the setup for observing {orbit}-orbit mixing of vector beam by a tilted aperture. P1 and P2, rotatable linear polarizers.}
		\label{Fig:Ex_Model}
	\end{figure}
	The Fourier spectrum of the field in the aperture plane can now be obtained using standard 2D Fast Fourier Transform (FFT) algorithms.
	\begin{eqnarray}
		\tilde{\boldsymbol{E}}(k_x,k_y) = \mathcal{FT}\left\{  \boldsymbol{E}(x,y,z=0) \right\} \label{eq_Fowd_FT}.
	\end{eqnarray} 
	Since our goal is to get the image in the plane perpendicular the path of the beam (in the $x'y'$ plane), we perform a rotation from un-primed to primed coordinate system in the $k$-space to get $\boldsymbol {\tilde{E}}(k_{x'},k_{y'})$. The transformation relations we use are given by,
	\begin{eqnarray}
		&k_{x'} = k_x\cos\theta_i - k_z\sin\theta_i , \quad k_{y'} =  k_y,&\notag \\ 
		&k_{z'} =  k_z\cos\theta_i + k_x\sin\theta_i. \label{eq_k_trans}&
	\end{eqnarray}
	Finally, we propagate the fields by a distance of $z'$ and calculate the field in the image plane, $\boldsymbol{E_{I}}$ by taking an inverse Fourier transform as follows,
	\begin{eqnarray}
		\boldsymbol{E_{I}}(r', z') = \mathcal{FT}^{-1} \left\{J(k_x',k_y') ~ \boldsymbol{E}(k_x',k_y') e^{ik_z' z'}\right\}. \label{eq_E_final}
	\end{eqnarray} 
	The Jacobian in eq. \ref{eq_E_final} is related to the rotation in $k$ space such that $dk_x \,dk_y = J(k_x',k_y')\, dk_x'\,dk_y'$ given by
	\begin{eqnarray}
		J(k_x',k_y') = \left|\cos\theta_i - \frac{k_x'}{k_z'}\sin\theta_i\right|.
	\end{eqnarray} 
	\par
	Note that while evaluating the forward transform in eq. \ref{eq_Fowd_FT}, we have taken evenly spaced $k_x$ values. However, from the nature of eq.  \ref{eq_k_trans}, it is clear that $k_{x'}$ has a non-linear dependence on $k_x$. Hence, a uniformly sampled $k_x$ values yields unevenly distributed $k_{x'}$ values. Hence, in order to employ standard fast Fourier algorithms to calculate the inverse Fourier transform, we first interpolate the data from $E(k_{x'},k_{y'})$ to calculate the field at uniformly spaced values of $k_{x'}$. Furthermore, the transformation causes the $k_{x'}$ values to fold on itself at the point where $k_z'$ goes to zero. In other words, we find that two distinct values of $k_x$ transform to the same value of $k_{x'}$, each with opposite signs for $k_{z'}$. We ignore contributions from all points where $k_z'<0$, since they cannot form an image at the destination plane. 
		\begin{figure}[t]
		
		\subfloat[\label{fig:0deg:a}]{
			\includegraphics[width=0.48\columnwidth]{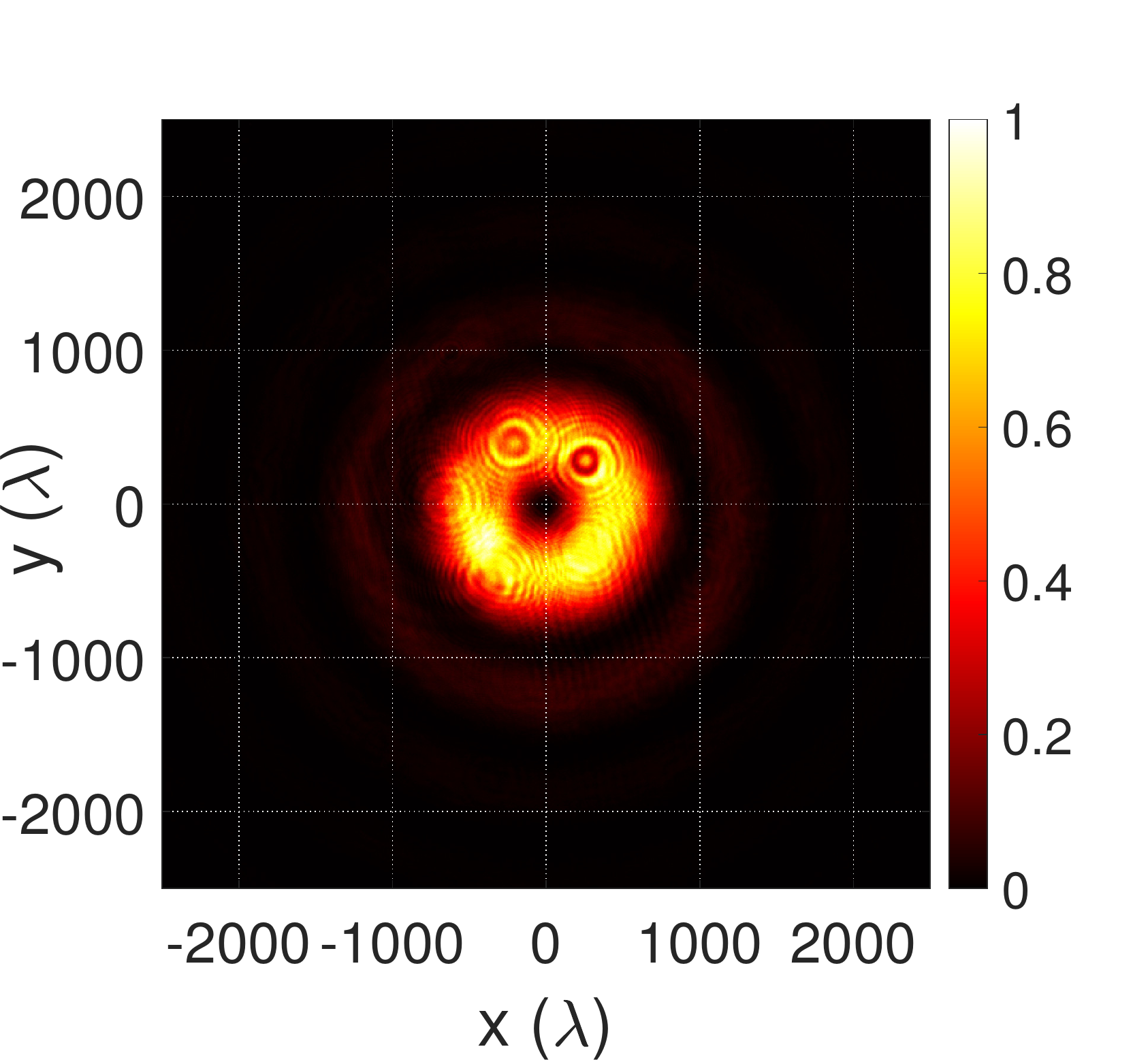}
		}\hspace*{\fill}%
		\subfloat[\label{fig:0deg:b}]{
			\includegraphics[width=0.48\columnwidth]{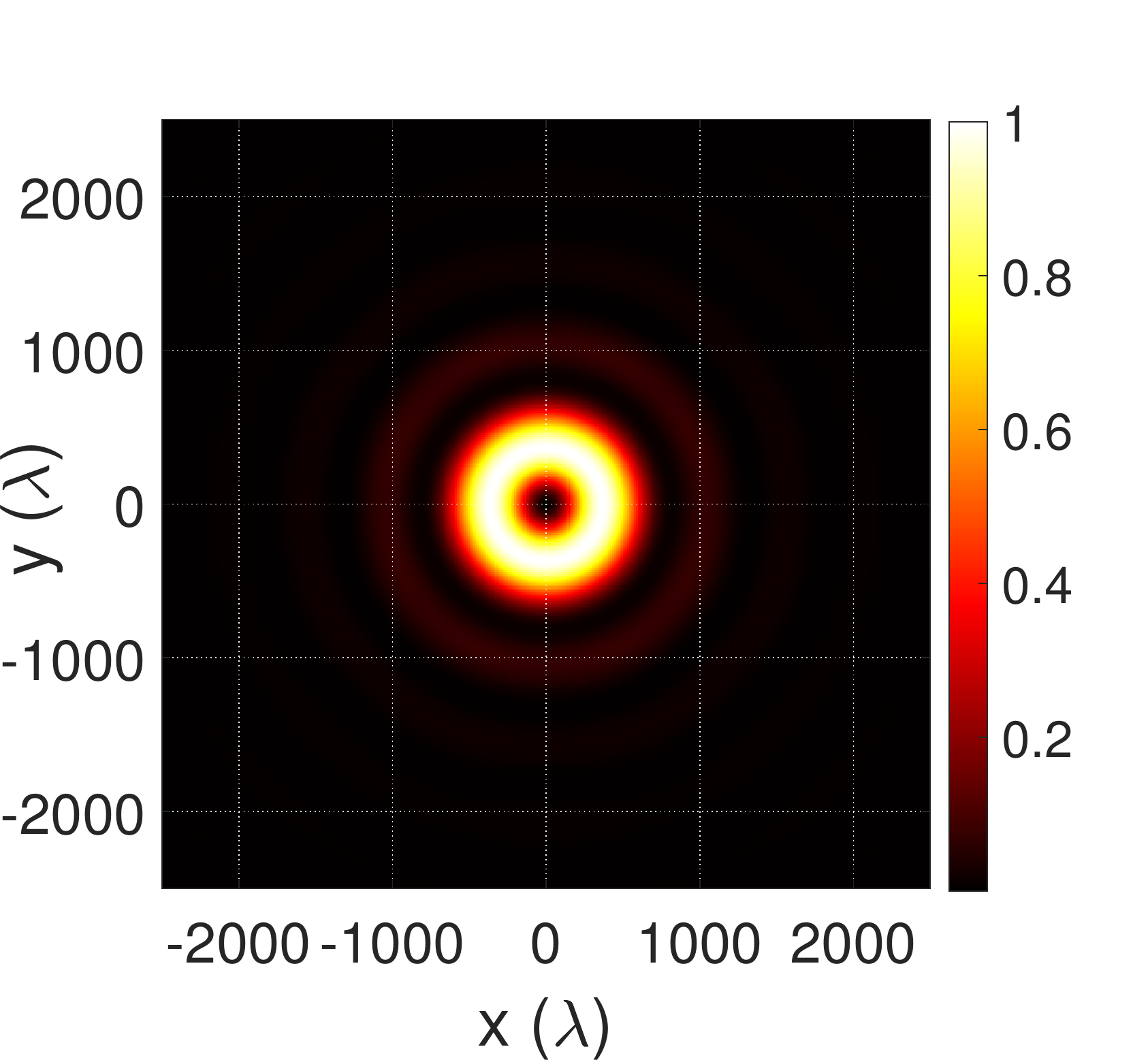}
		}
		
		\medskip
		
		\subfloat[\label{fig:0deg:c}]{
			\includegraphics[width=0.48\columnwidth]{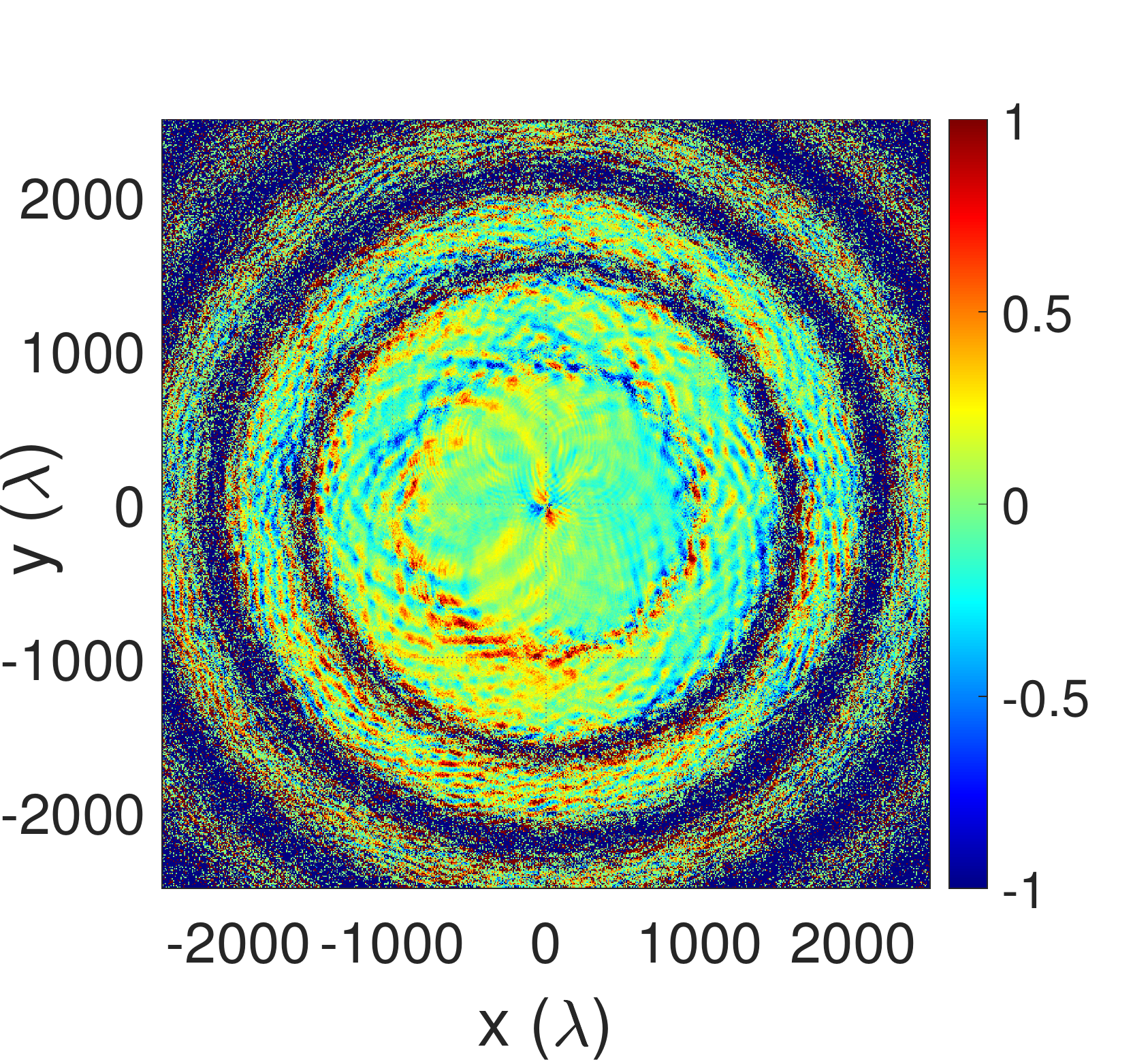}
		}\hspace*{\fill}
		\subfloat[\label{fig:0deg:d}]{
			\includegraphics[width=0.48\columnwidth]{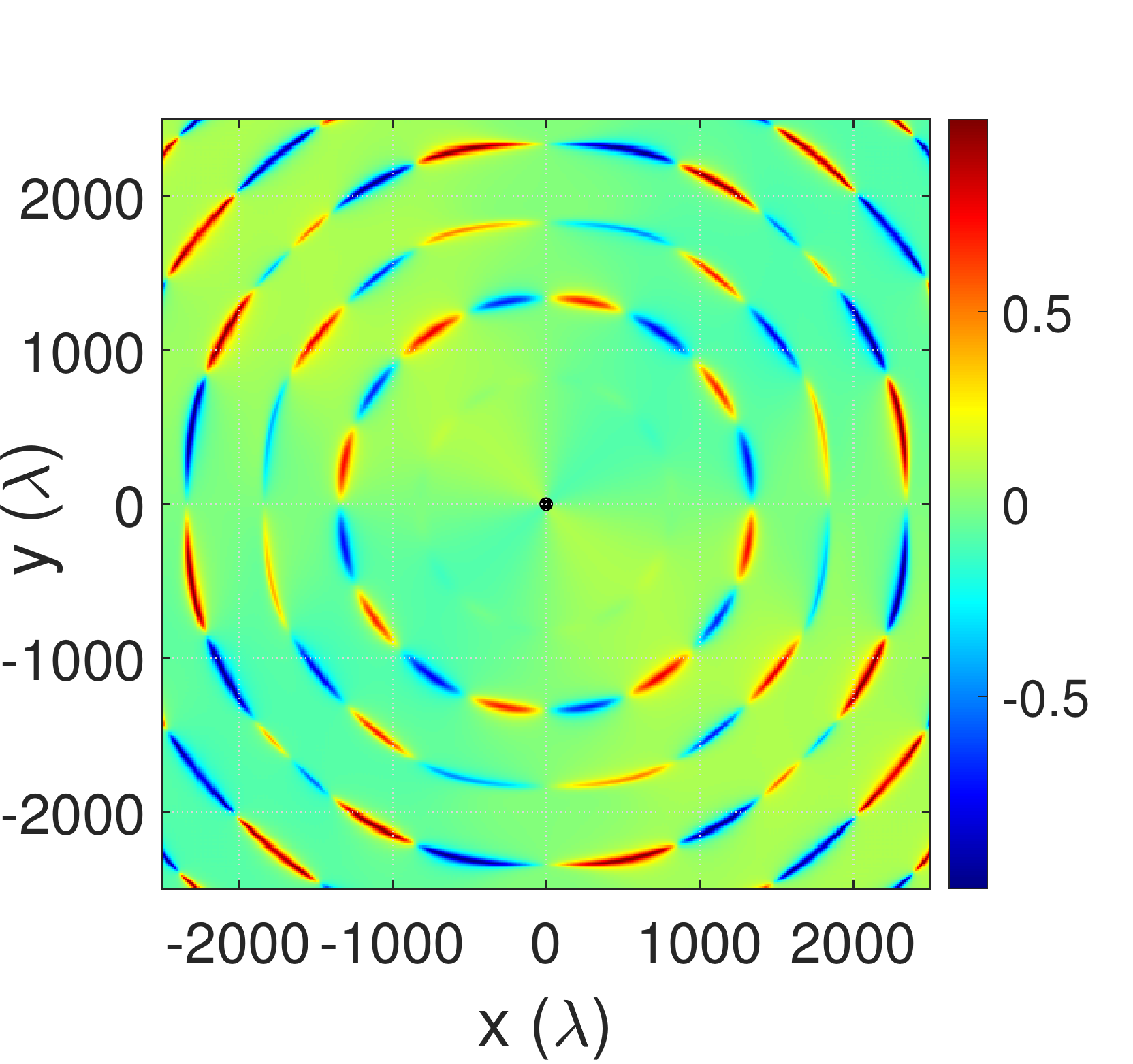}
		}	
		\caption{(a) Experimentally measured diffraction pattern from an aperture tilted at an angle $\theta_i = 0^0$. (b) Theoretical diffraction pattern from an aperture tilted at an angle $\theta_i = 0^0$ ($x_0 = 0$, $y_0  =0$). (c) Experimentally measured V/I, where V is the difference between left and right polarized light intensities. (d) Theoretical V/I.  The CG of the intensity pattern for LCP  and RCP components  are superposed using white and black dots, which coincide for this case.} \label{fig:0deg}
	\end{figure}
	\subsection{Experimental realization \label{Sec:Experimental_realization}}
	As mentioned in the introduction, the goal of this paper is to study orbit-orbit mixing both theoretically and experimentally. The above theoretical formulation was implemented in order to explain the observed results of the experiment we performed to study orbit-orbit coupling. A schematic of the experimental arrangement is shown in Fig. \ref{Fig:Ex_Model}. Fundamental Gaussian mode of a $632.8$ nm line of a He-Ne laser (HNL225R – He-Ne Laser, 632.8 nm, 22.5 mW, Random) having random polarization is used to produce the vector beam. The radially polarized vector beam was generated from the fundamental Gaussian mode by using a combination of a linear polarizer, P1 (GTH10M, Thorlabs, USA) and S-waveplate (RPC-515-06-46, Altechna). The radially polarized beam is passed through a tilted aperture (P300D, Thorlabs, USA) of diameter 300 $\mu$m. The diffraction pattern is then passed through a combination of quarter waveplate (WPQ10M-633, Thorlabs, USA) and linear polarizer, P2. The spatial variation of Stokes parameters [\textit{I Q U V}]$^T$ of the diffraction pattern are determined by the quarter waveplate and the linear polarizer combination. The measurements were performed for a range of tilt angle of the aperture ($\theta_i = 0^{\circ}, 7^{\circ}, 14^{\circ}, 40^{\circ}$). The polarization resolved diffraction pattern are imaged into a CCD camera (1024 $\times$ 768 square pixels, pixel dimension 4.65 $\mu m$, Thorlabs, USA)  at a distance of $z' = 15$ cm. The image collected by the CCD is compared with the theoretical simulations.
	\begin{figure}[t]
		
		\subfloat[\label{fig:7deg:a}]{
			\includegraphics[width=0.48\columnwidth]{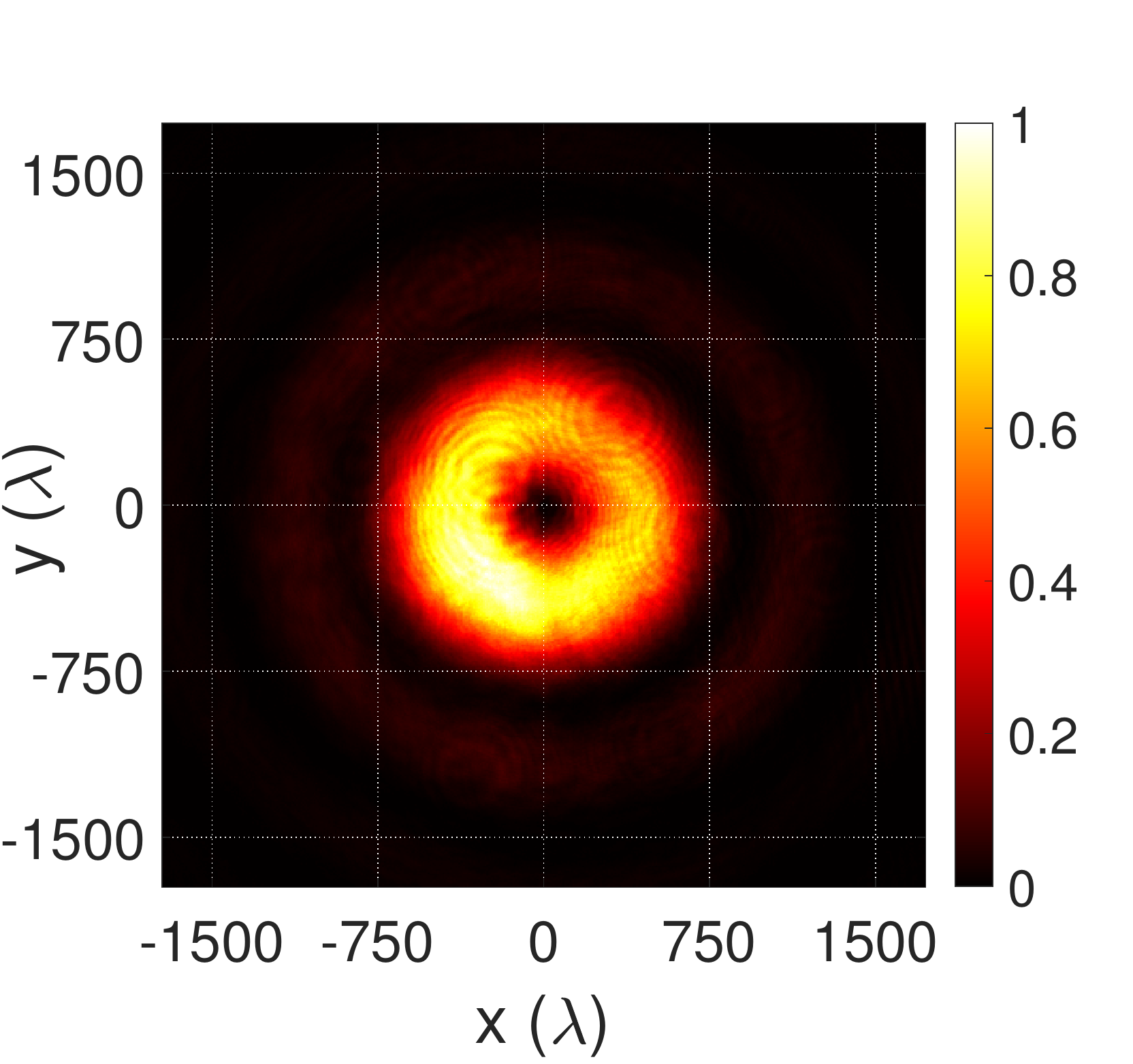}
		}\hspace*{\fill}%
		\subfloat[\label{fig:7deg:b}]{
			\includegraphics[width=0.48\columnwidth]{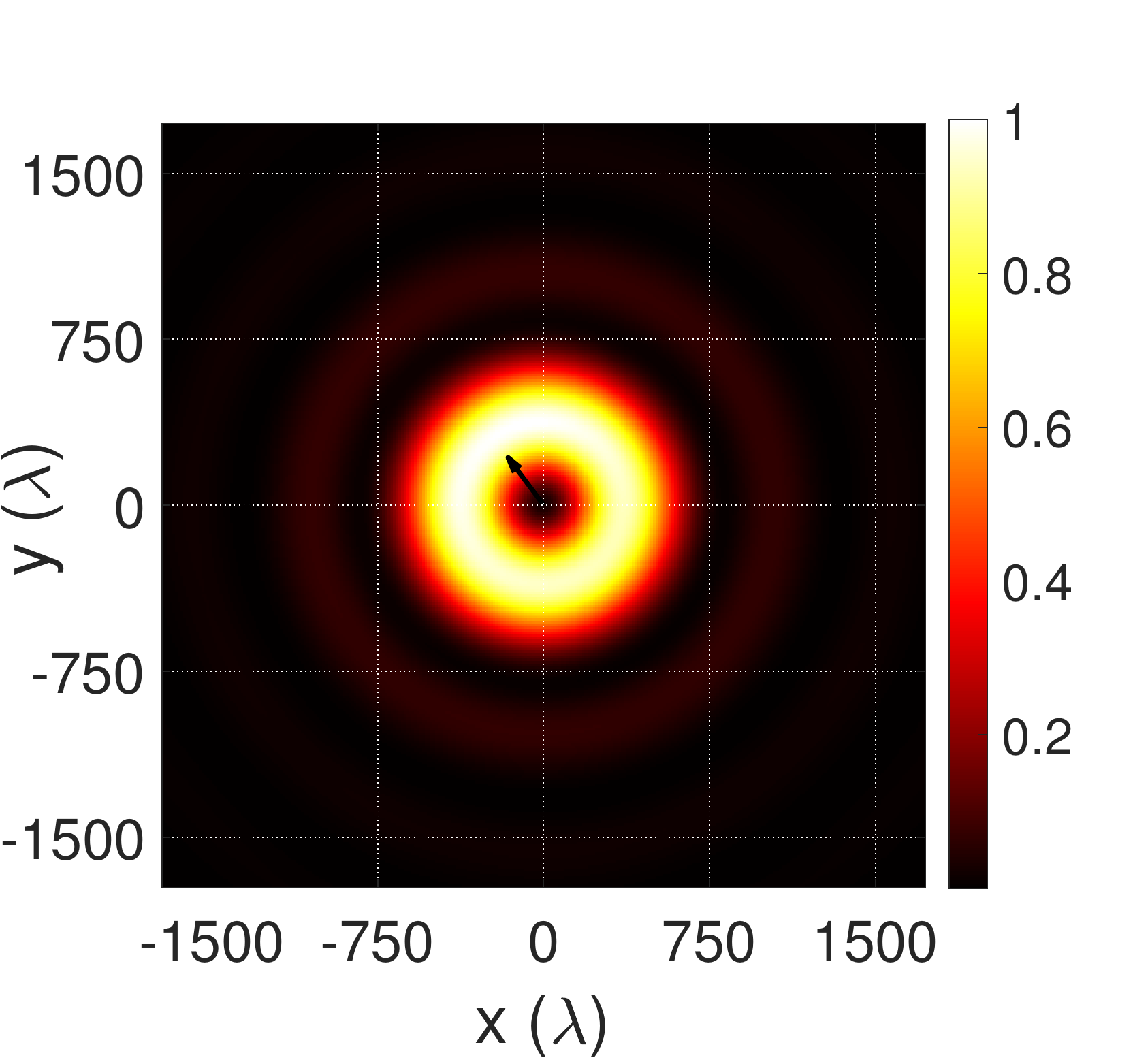}
		}
		
		\medskip
		
		\subfloat[\label{fig:7deg:c}]{
			\includegraphics[width=0.48\columnwidth]{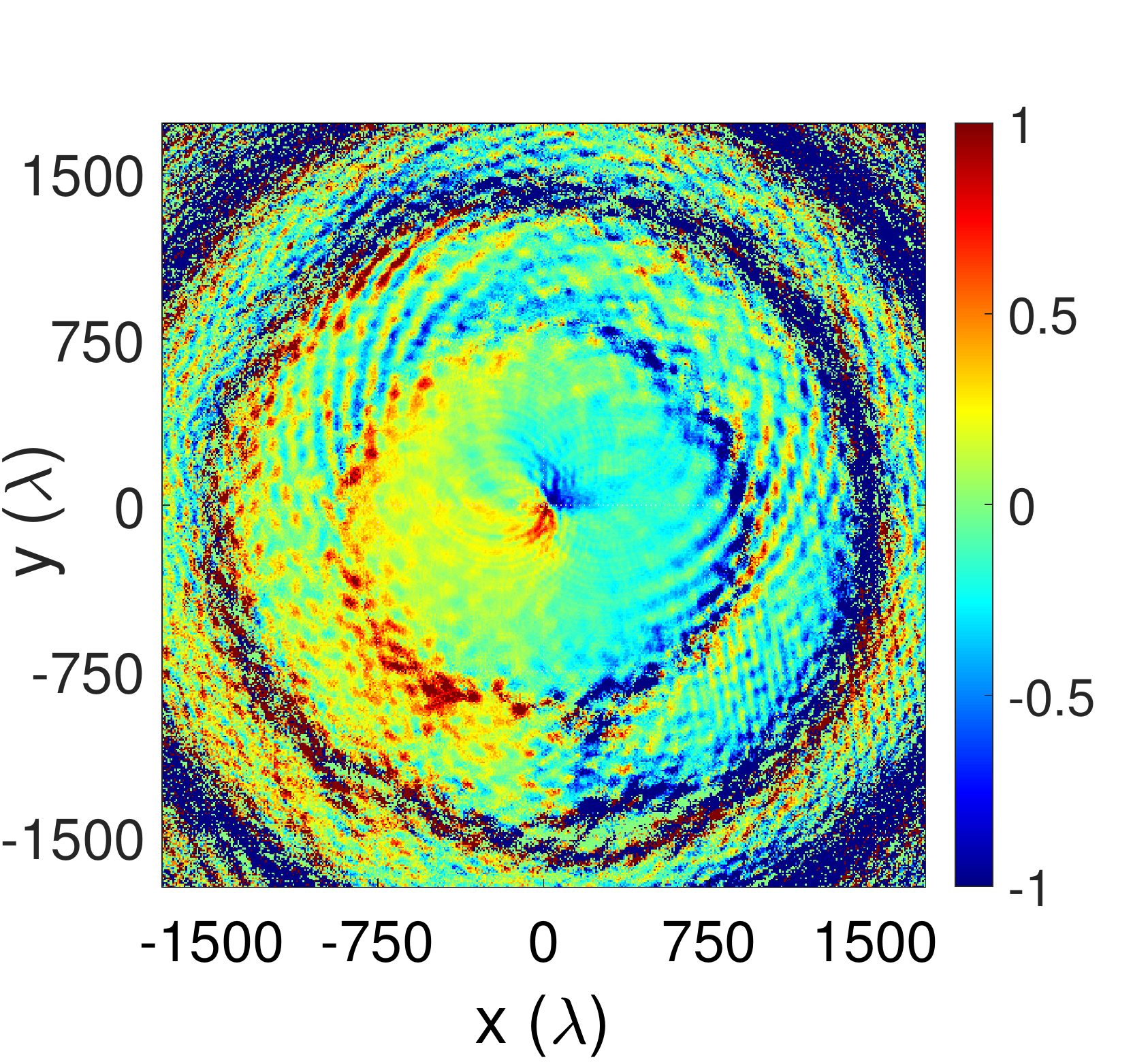}
		}\hspace*{\fill} 
		\subfloat[\label{fig:7deg:d}]{%
			\includegraphics[width=0.48\columnwidth]{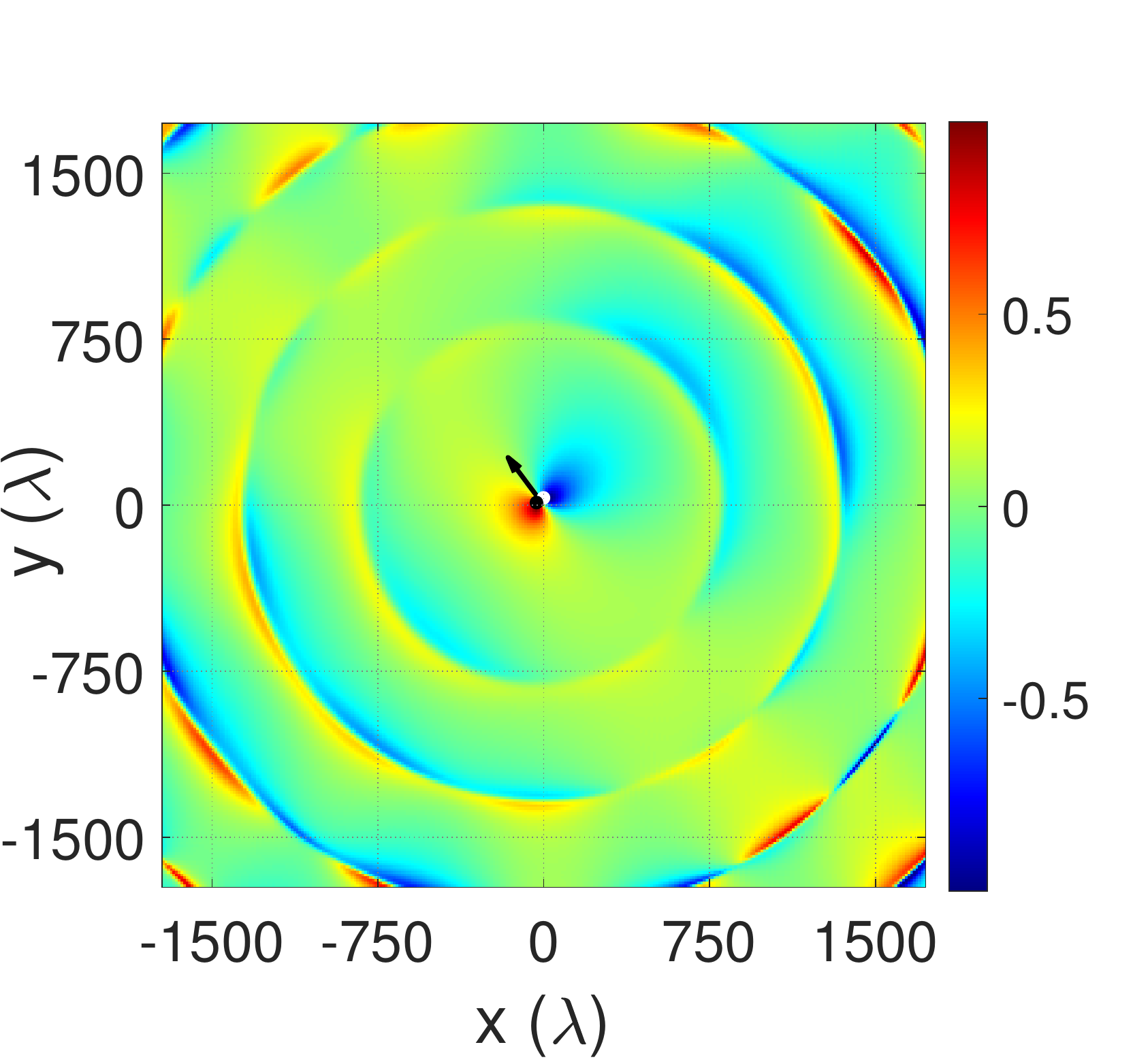}%
		}	 
		\caption{(a) Experimentally measured diffraction pattern from an aperture tilted at an angle $\theta_i = 7^0$. (b) Theoretical diffraction pattern from an aperture tilted at an angle $\theta_i = 7^0$ ($x_0\sim -9\lambda$, $y_0=12\lambda$). (c) Experimentally measured V/I, where V is the difference between left and right polarized light intensities. (d) Theoretical V/I. The CG of the intensity pattern for LCP  and RCP components  is superposed using white and black dots.} \label{fig:7deg}
	\end{figure}
	\begin{figure}[t]
		
		\subfloat[\label{fig:14deg:a}]{
			\includegraphics[width=0.48\columnwidth]{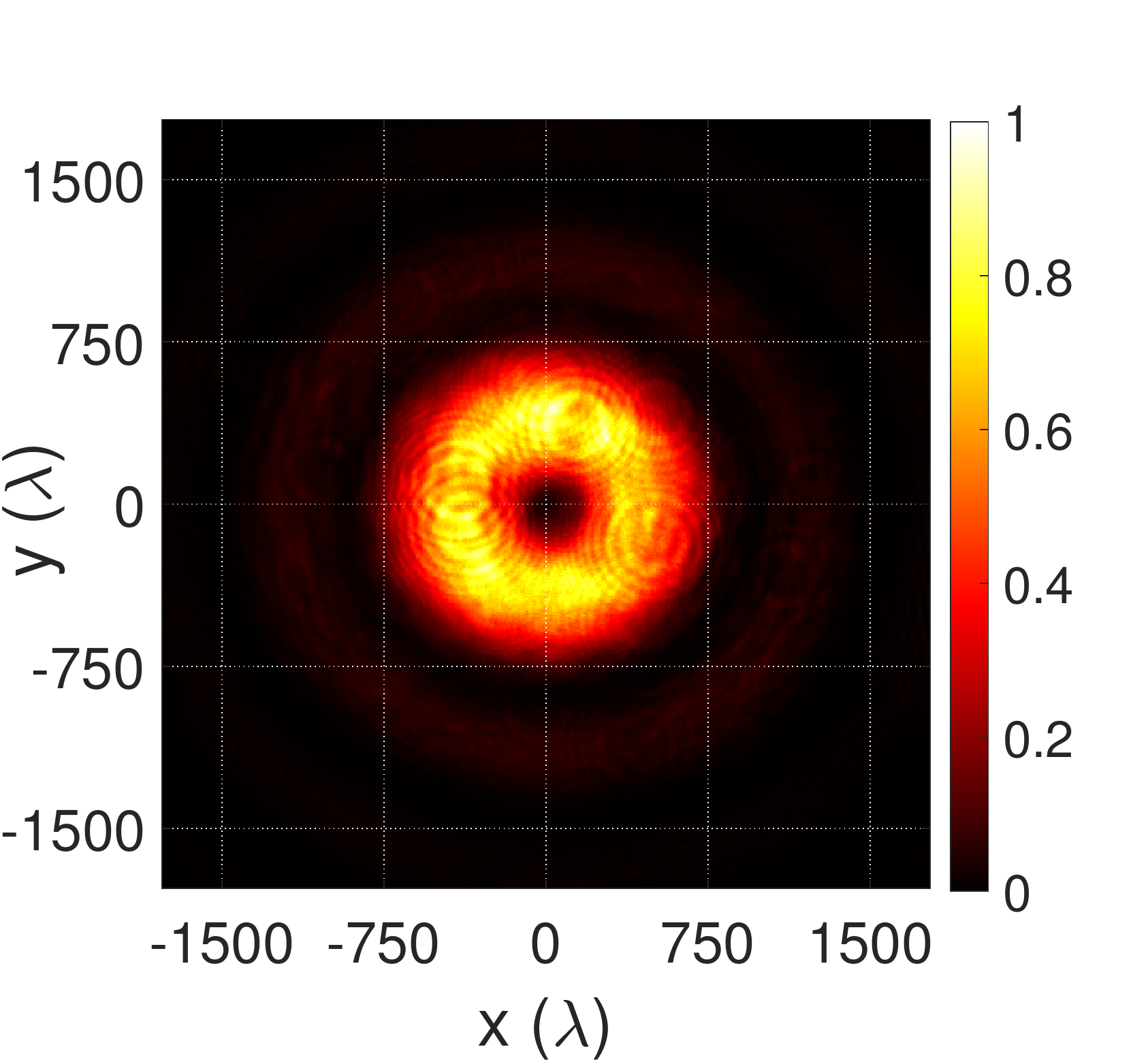}
		}\hspace*{\fill}%
		\subfloat[\label{fig:14deg:b}]{
			\includegraphics[width=0.48\columnwidth]{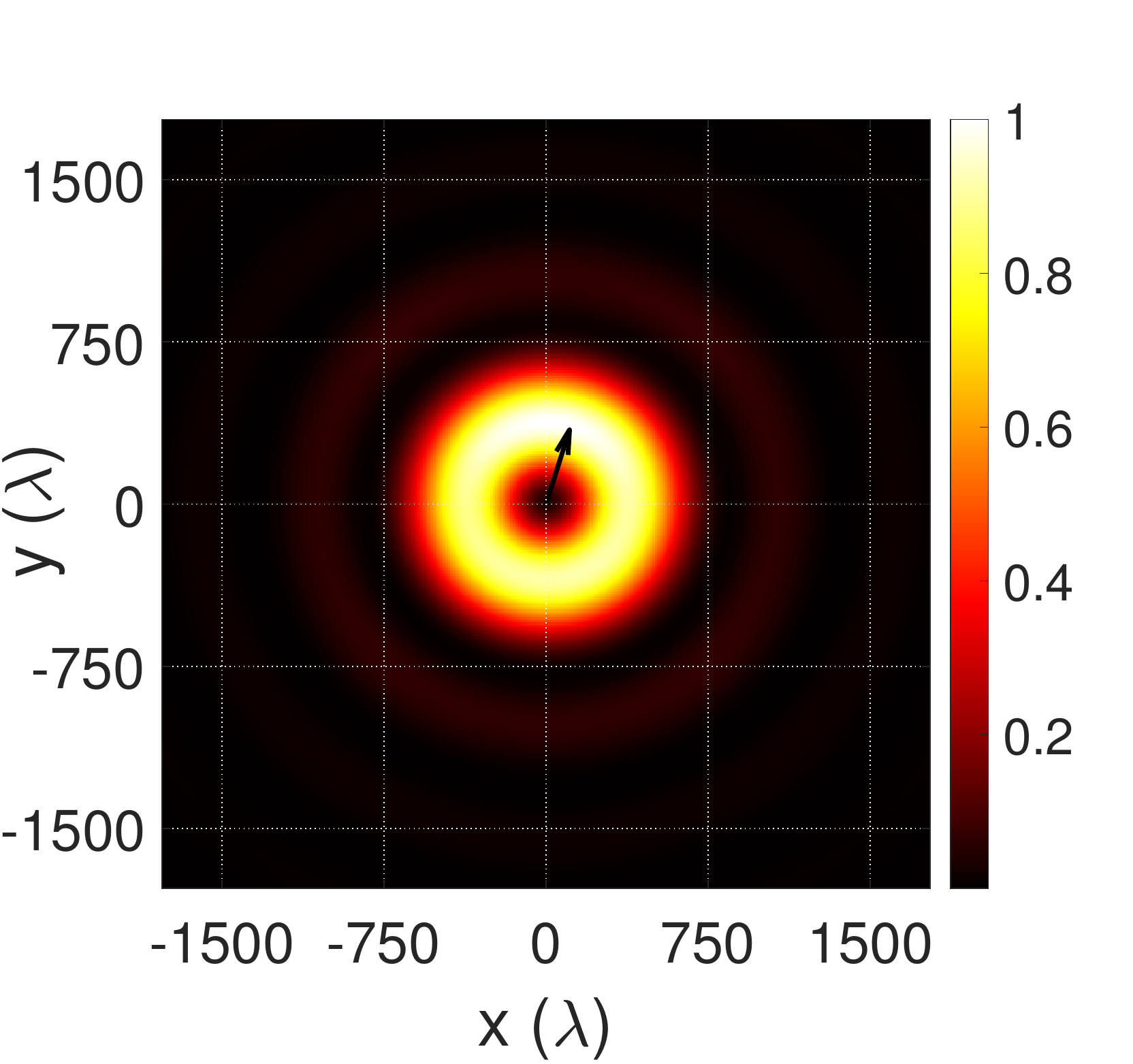}
		}
		
		\medskip
		
		\subfloat[\label{fig:14deg:c}]{
			\includegraphics[width=0.48\columnwidth]{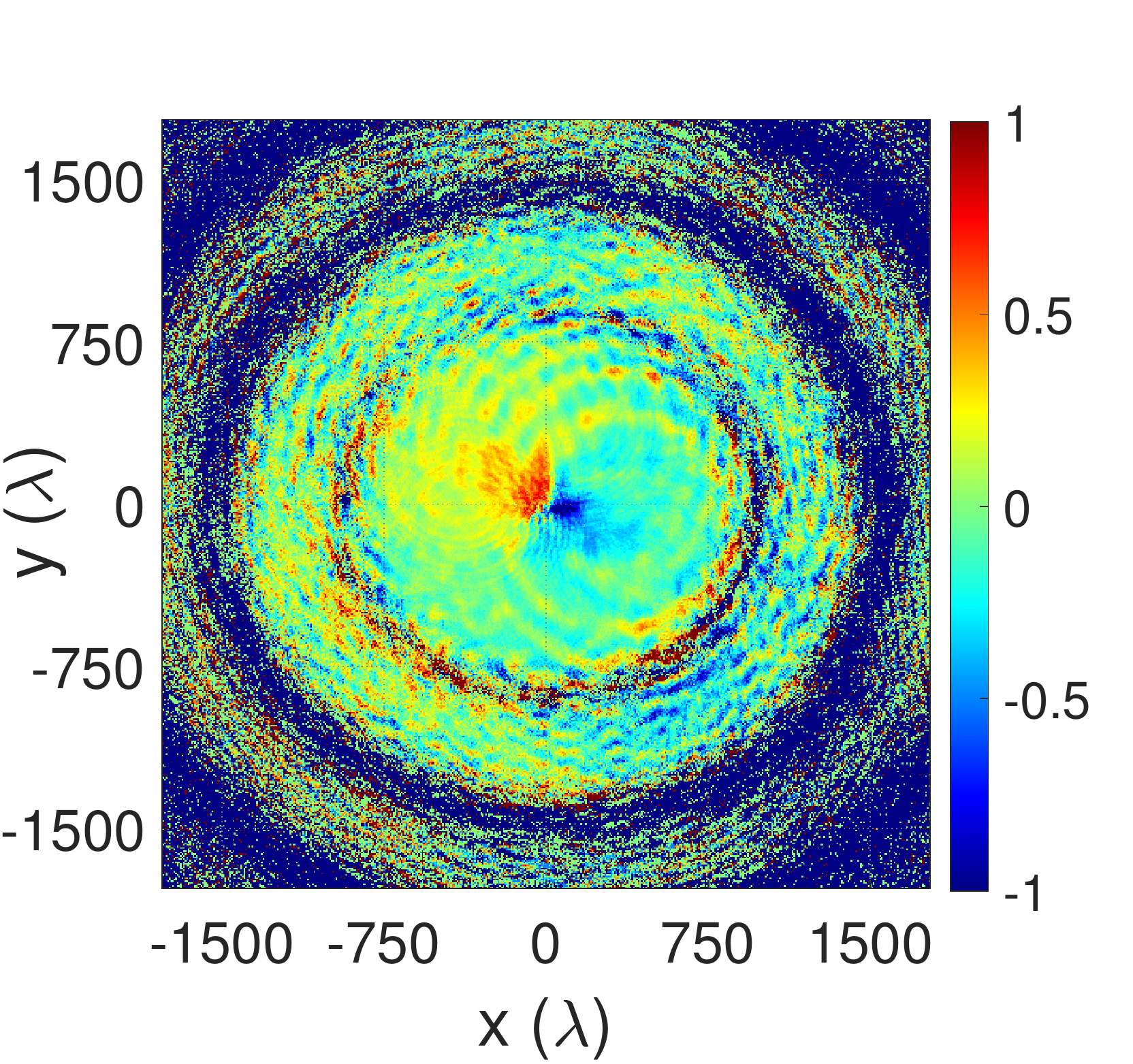}
		}\hspace*{\fill}
		\subfloat[\label{fig:14deg:d}]{%
			\includegraphics[width=0.48\columnwidth]{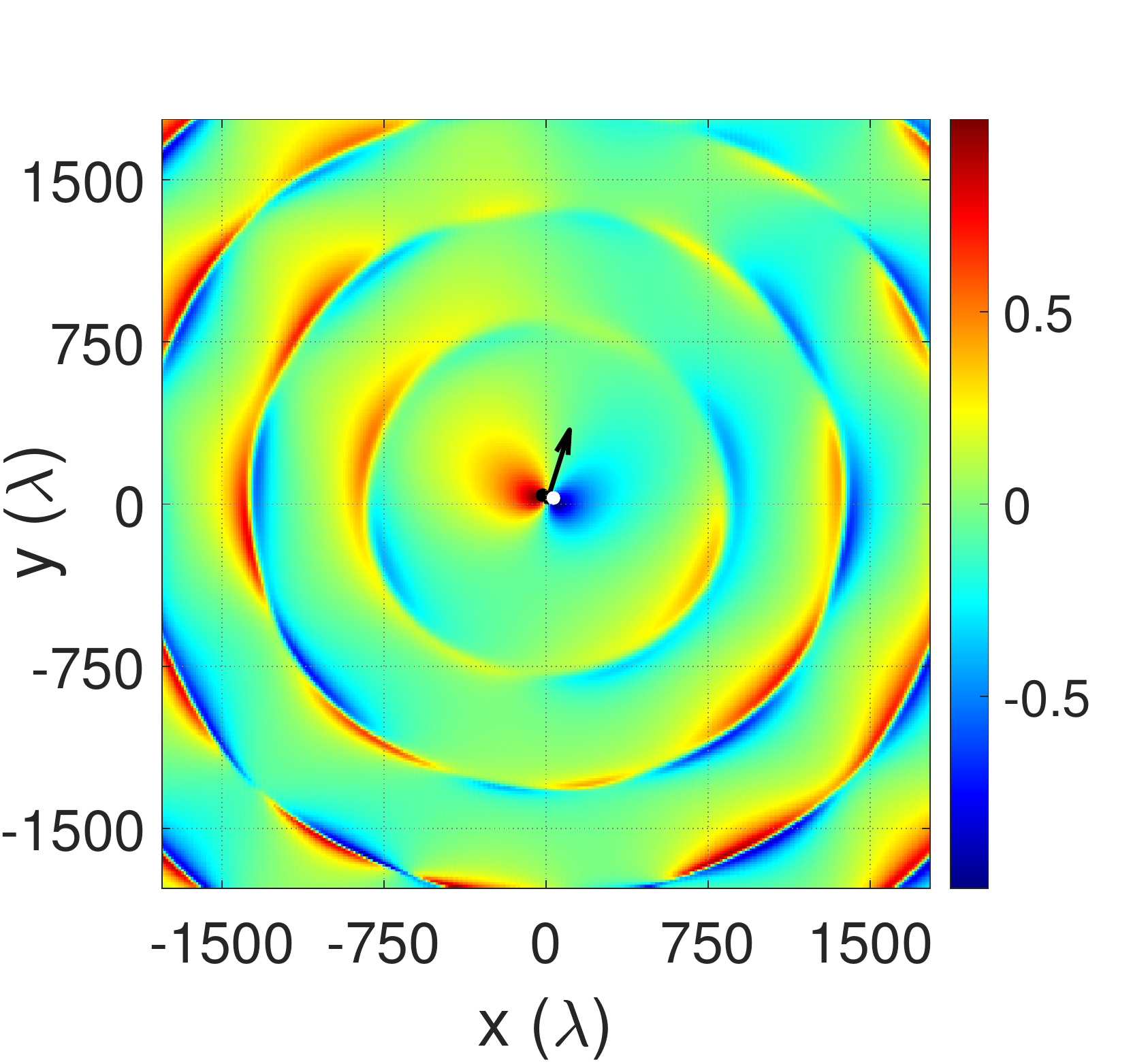}%
		}				
		\caption{(a) Experimentally measured diffraction pattern from an aperture tilted at an angle $\theta_i = 14^0$. (b) Theoretical diffraction pattern from an aperture tilted at an angle $\theta_i = 14^0$ ($x_0  \sim 6\lambda$ and $y_0 \sim  19.1 \lambda$). (c) Experimentally measured V/I, where V is the difference between left and right polarized light intensities. (d) Theoretical V/I.  The CG of the intensity pattern for LCP  and RCP components are superposed using white and black dots.} \label{fig:14deg}
	\end{figure}
	\begin{figure}[t]
	
	\subfloat[\label{fig:40deg:a}]{
		\includegraphics[width=0.48\columnwidth]{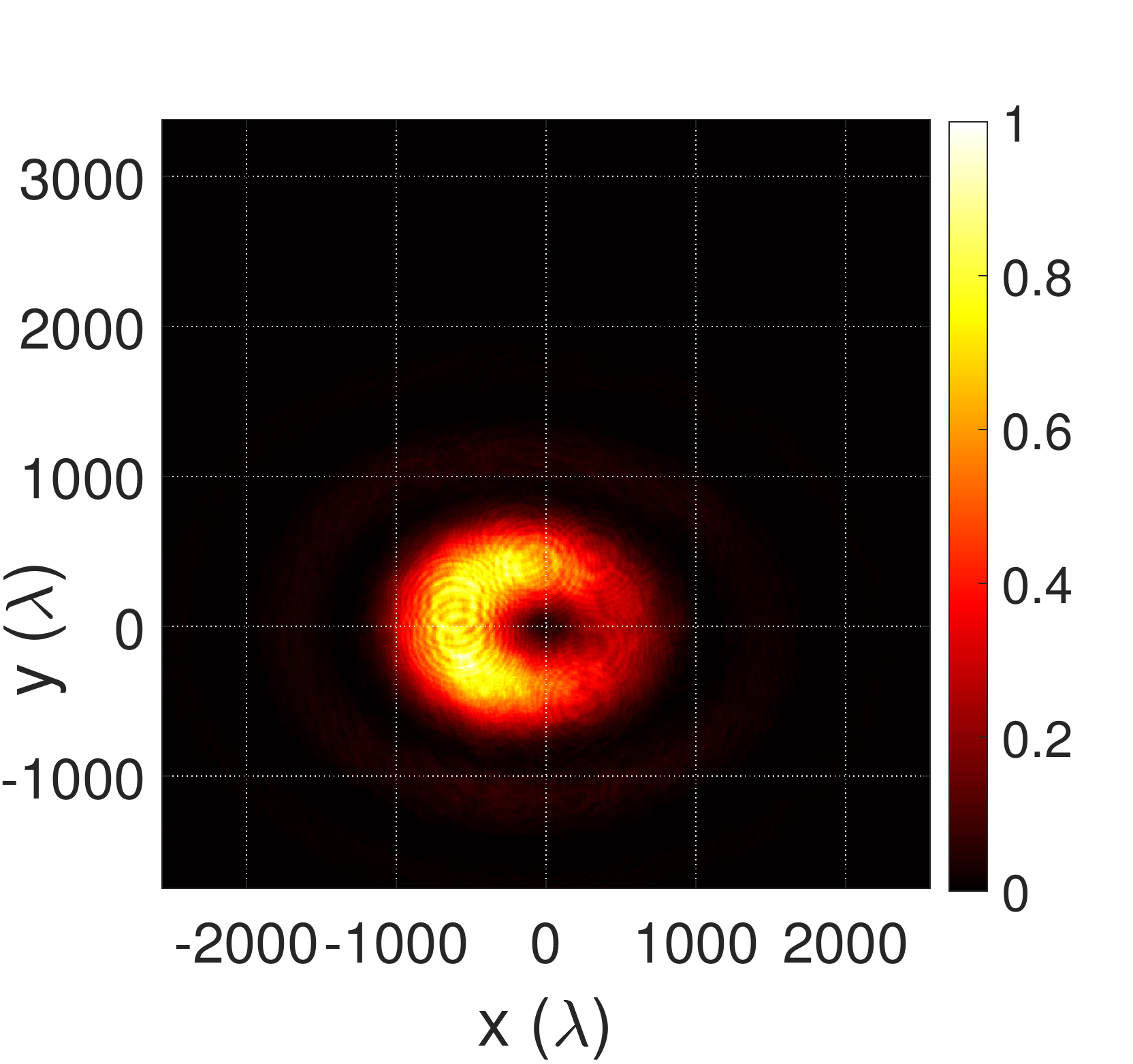}
	}\hspace*{\fill}%
	\subfloat[\label{fig:40deg:b}]{
		\includegraphics[width=0.48\columnwidth]{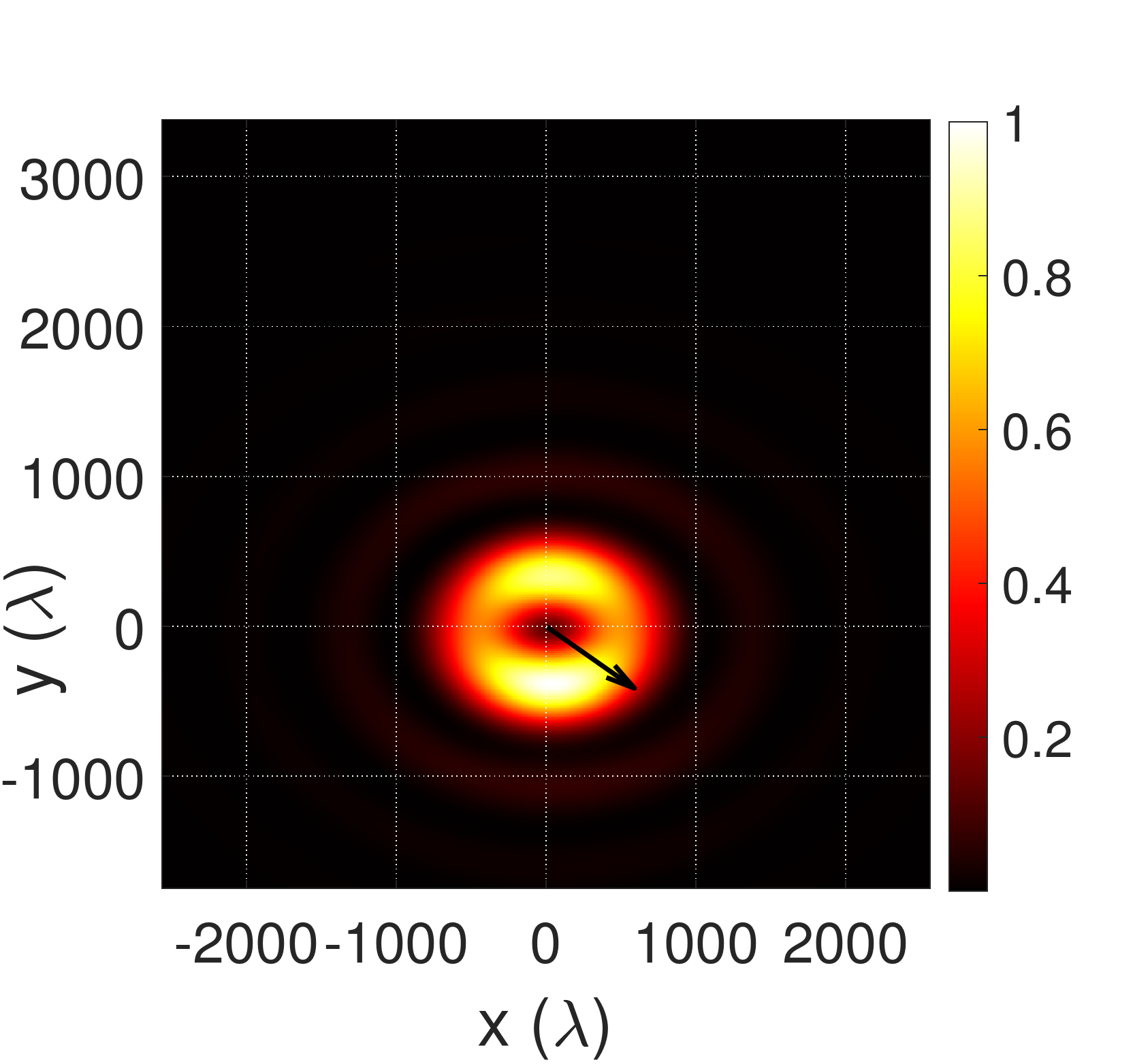}
	}
	
	\medskip
	
	\subfloat[\label{fig:40deg:c}]{
		\includegraphics[width=0.48\columnwidth]{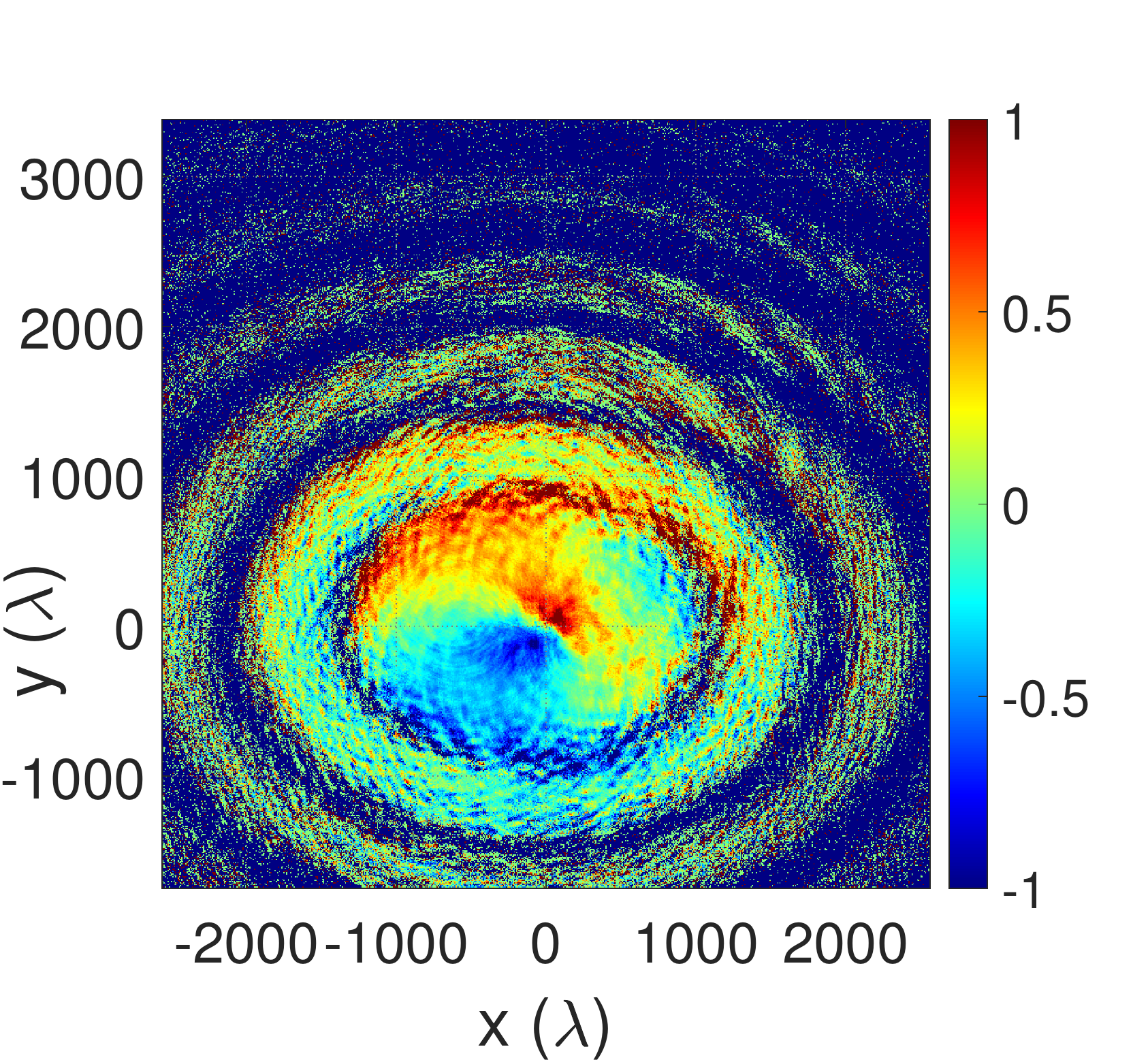}
	}\hspace*{\fill}
	\subfloat[\label{fig:40deg:d}]{%
		\includegraphics[width=0.48\columnwidth]{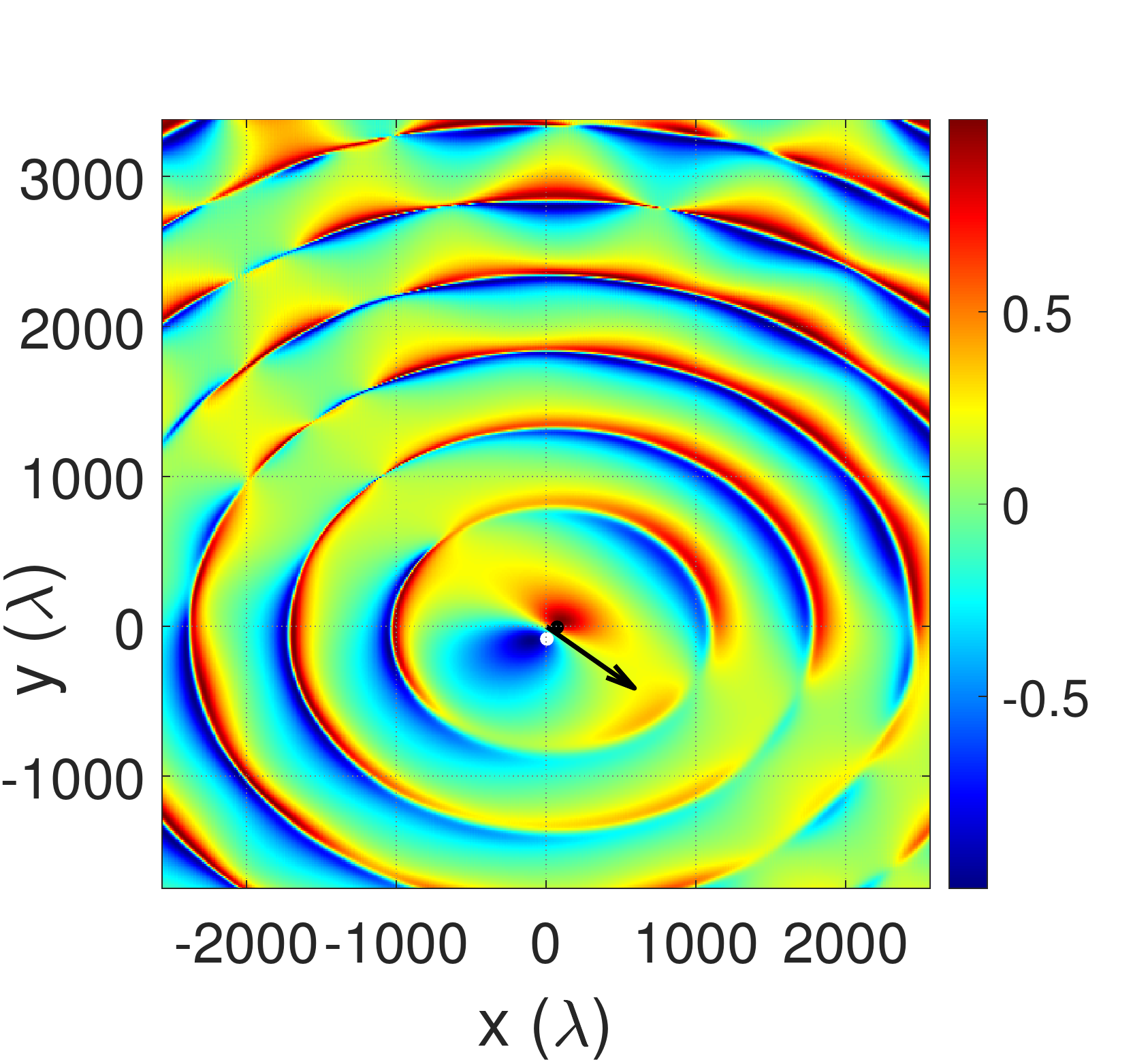}%
	}		
	\caption{(a) Experimentally measured diffraction pattern from an aperture tilted at an angle $\theta_i = 40^0$. (b) Theoretical diffraction pattern from an aperture tilted at an angle $\theta_i = 40^0$ ($x_0 \sim 32.8\lambda $ and $y_0 \sim -22.9\lambda$). (c) Experimentally measured V/I, where V is the difference between left and right polarized light intensities. (d) Theoretical V/I.  The CG of the intensity pattern for LCP  and RCP components are superposed using white and black dots.} \label{fig:40deg}
\end{figure}
	\section{\label{sec:level3}Experimental and numerical results and discussions}
	\subsection{\label{sec:level3.1} Comparison between experimentally measured and simulation results}	
	We use a nearly radially polarized vector beam to model the experimental beam we use. By examining the Stokes mapping of the bare vector beam used in the experiment (not shown), a small eccentricity has been included in our numerical calculations to account for the experimentally unavoidable errors. {In particular, in order to have proper correspondence between theory and experiment we first studied the bare experimental beam and chose $\delta$ and $\theta$ to yield the theoretical intensity distribution that best matches the experimental pattern. These values are used for all the Stokes images. The same approach is used to arrive at the value of the shift.} 
	This non-zero eccentricity for the polarization ellipse formed by $E_{x'}$ and $E_{y'}$ is modeled using  $\theta=2^0$ and $\delta=5^0$ in eq. \ref{eq_VB}.
	\par {In order to have a clear idea about the individual effects of shift and tilt on the displacement of the beam we calculated the center of gravity (CG) of the beams (first moments) only theoretically since experimental realization of a given set of values for tilt and shift is practically impossible. Moreover, so as to have a parallel with Stokes images for degree of circular polarization, we projected the electric field $\vec E$ onto the circular basis, $\vec E = E_{LCP} \,\hat e_{LCP} + E_{RCP}\, \hat e_{RCP}$, where $E_{LCP,RCP} = (E_x \mp iE_y)/\sqrt{2}$ . The first moment of the intensity distribution in the circular basis can then be calculated using
	\begin{eqnarray}
		<x_{LCP,RCP}> = \frac{\int x |E_{LCP,RCP}|^2 dx\,dy}{\int  |E_{LCP,RCP}|^2 dx\,dy}\\
		<y_{LCP,RCP}> = \frac{\int y |E_{LCP,RCP}|^2 dx\,dy}{\int |E_{LCP,RCP}|^2 dx\,dy}
	\end{eqnarray}}
	\par
	Figs. \ref{fig:0deg:a}, \ref{fig:7deg:a}, \ref{fig:14deg:a} and \ref{fig:40deg:a} show the experimentally measured intensity plots for a vector beam diffracted from a tilted aperture with a diameter of $300$ $\mu m$ for increasing angles. The corresponding  simulated intensity plots are shown in Figs. \ref{fig:0deg:b}, \ref{fig:7deg:b}, \ref{fig:14deg:b} and \ref{fig:40deg:b}. Figs. \ref{fig:0deg:c}, \ref{fig:7deg:c}, \ref{fig:14deg:c} and \ref{fig:40deg:c} show the experimentally measured Stokes parameter plots for the corresponding intensity. However, due to the null intensity at the center of the vector beam (resulting in the inaccuracy of pinning down the center), there could be a small error (of the order of a maximum of 40$\lambda$ or 25 $\mu$m ) in assessing the distance between the centers of the vector beam and the aperture. We thus include a small shift between the two centers in our numerical calculations, shown in Figs. \ref{fig:0deg:d}, \ref{fig:7deg:d}, \ref{fig:14deg:d} and \ref{fig:40deg:d}. The arrow in the theoretical plots show the direction of the shifts. Surprisingly, this small shift between the two centers causes a sufficiently large splittings between the  left and right circularly polarized lights in the Stokes plot, as can be seen in these figures. We see that with increasing split between the centers, the left and right circularly polarized lights separates out in the direction perpendicular to the shift. The remarkable agreement between the experimental and numerical calculations leads us to conclude that the measured splitting is a result of the shift between the two centers. It should be noted that though the intensity plots for the angles $\theta_i = 0^0$, $7^0$ and $14^0$ seems to be comparable with the numerical results, the one for $\theta_i=40^0$ seems to be different. This could be due to the slight asymmetry in the bare beam used in the experimental setup.
	\begin{figure}[t]
		\subfloat[\label{fig:V_w_w_out_shift:a}]{
			\includegraphics[width=0.48\columnwidth]{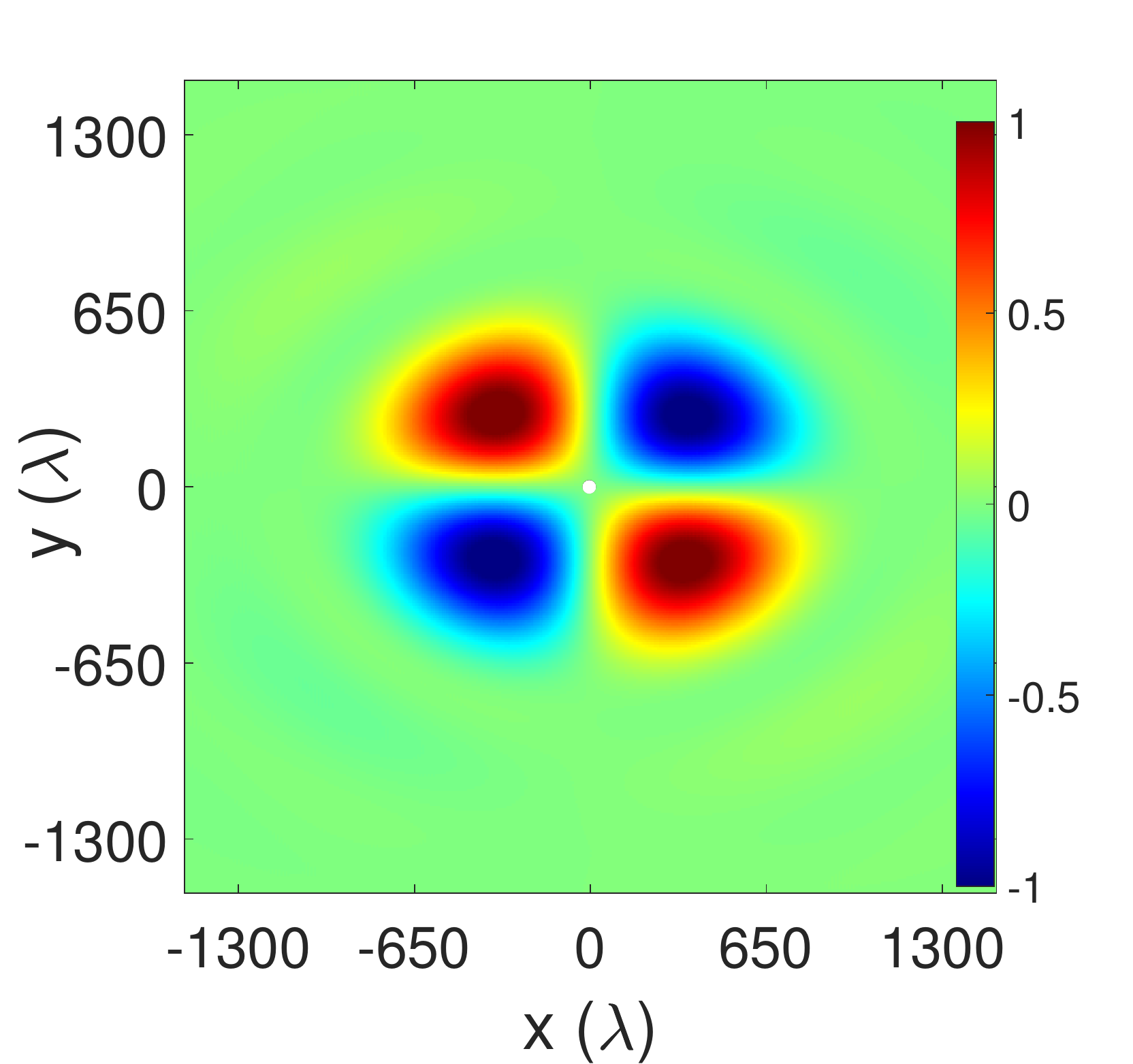}
		}\hspace*{\fill}%
		\subfloat[\label{fig:V_w_w_out_shift:b}]{
			\includegraphics[width=0.48\columnwidth]{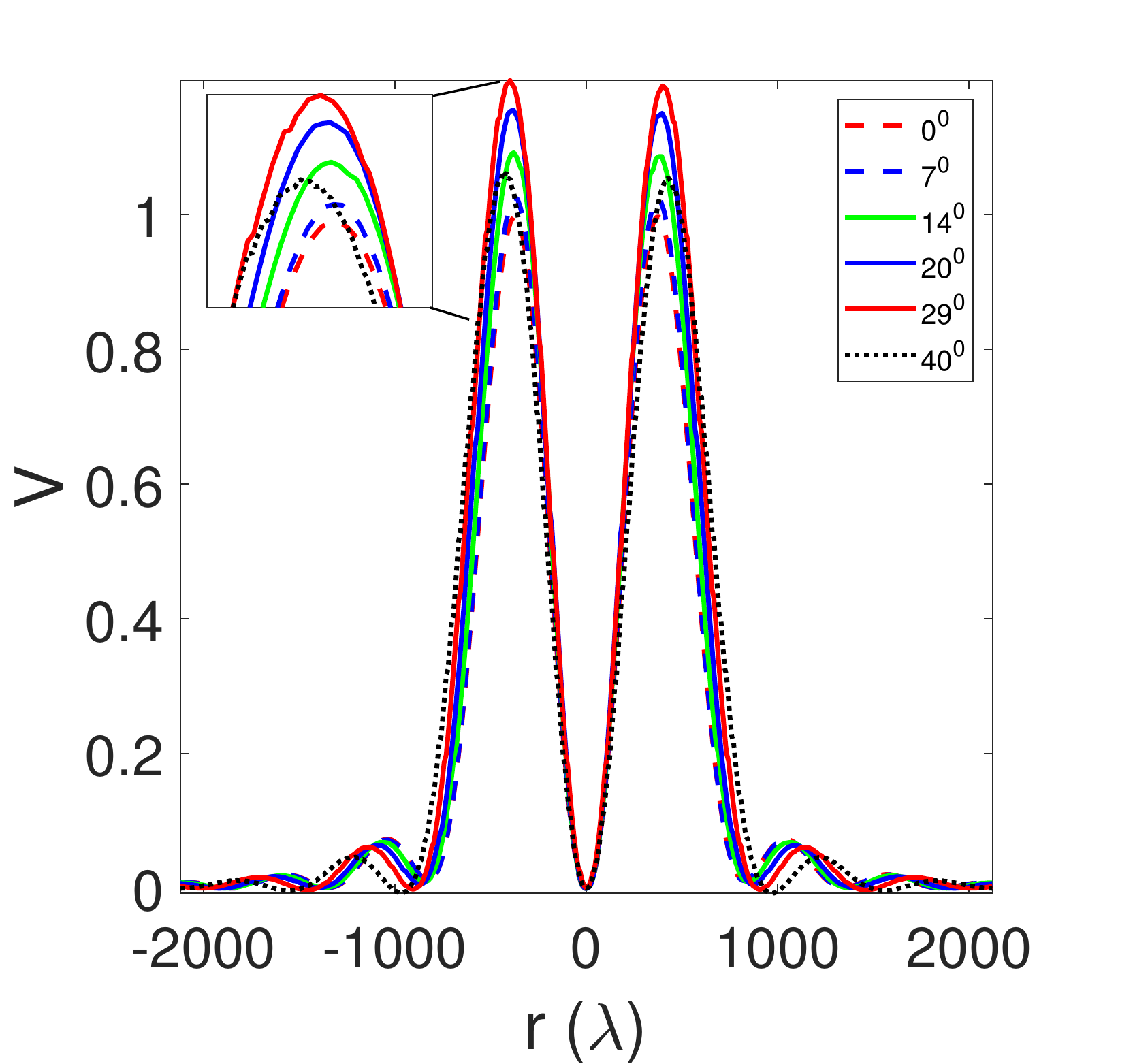}
		}
		
		\medskip
		
		\subfloat[\label{fig:V_w_w_out_shift:c}]{
			\includegraphics[width=0.48\columnwidth]{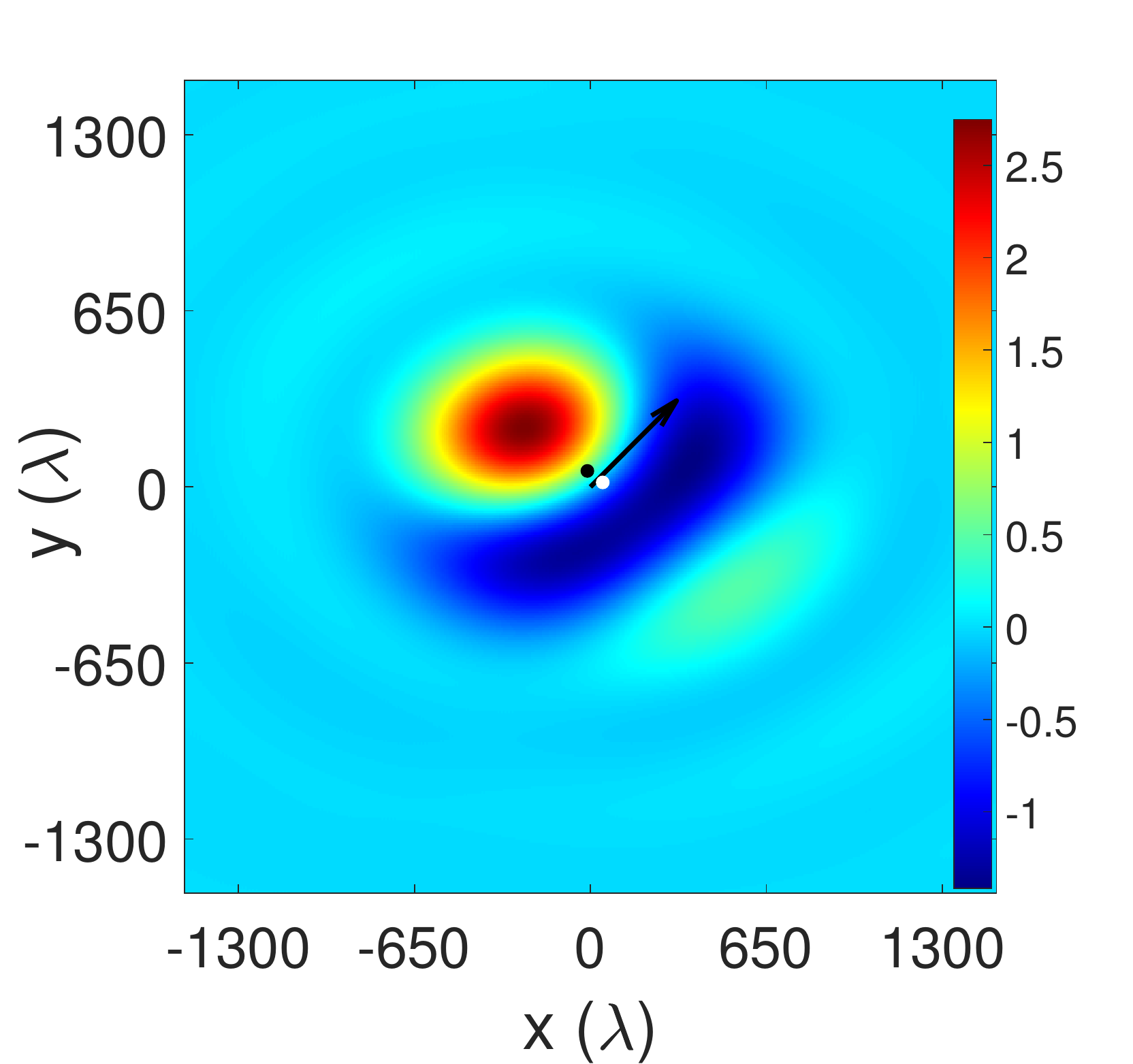}
		}\hspace*{\fill}
		\subfloat[\label{fig:V_w_w_out_shift:d}]{%
			\includegraphics[width=0.48\columnwidth]{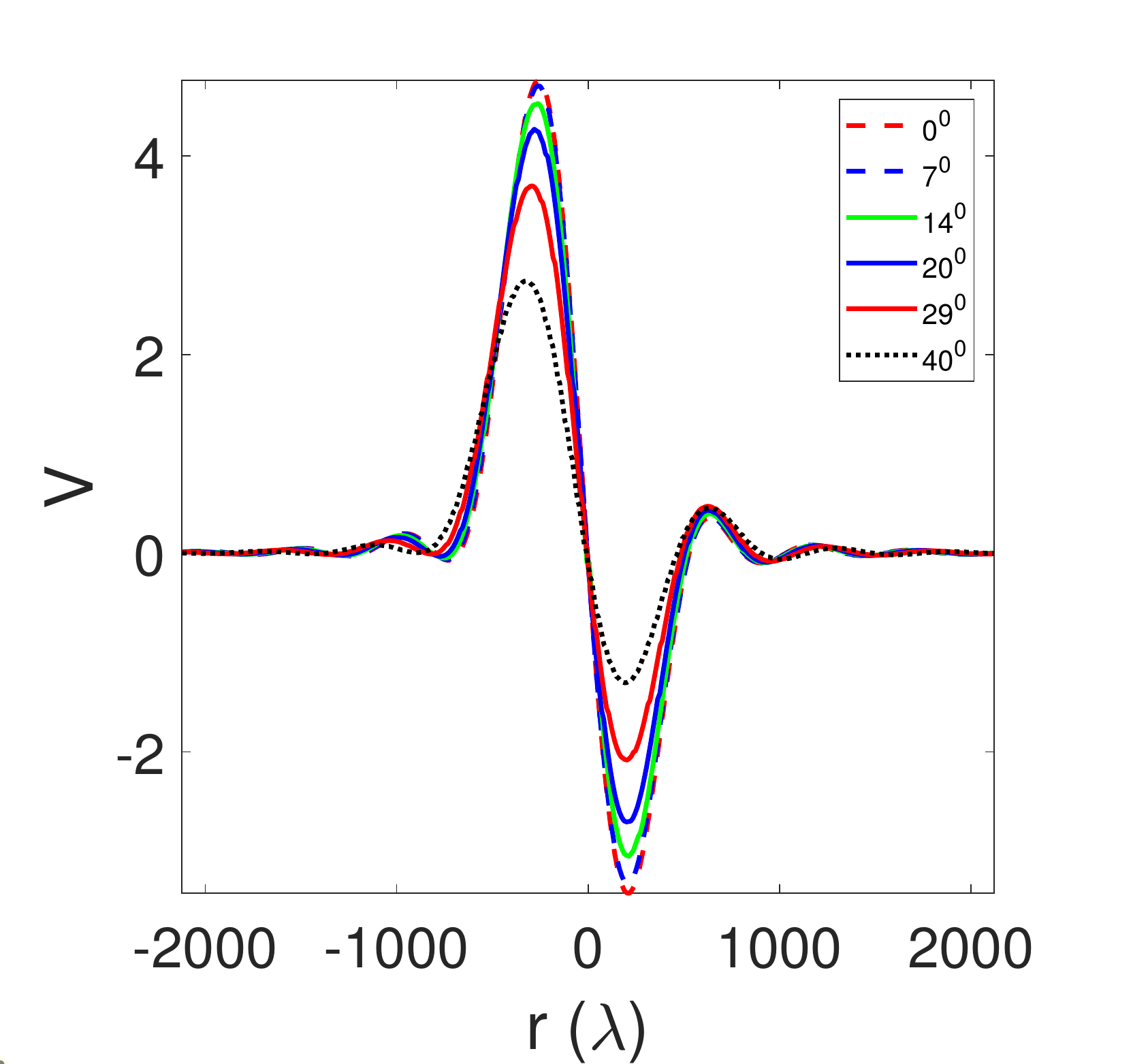}%
		}	 			
		\caption{(a) Numerically calculated Stokes map for a perfectly aligned aperture tilted at an angle $\theta_i=40^0$. (b) Variation of the Stokes parameter $V$ in the direction $(\hat{x}-\hat{y})/\sqrt 2$ for different angles of tilt $\theta_i$ for a perfectly aligned aperture. (c) Numerically calculated Stokes map for an off-axis aperture tilted at an angle $\theta_i=40^0$ and a shift of $x_0=y_0=17.67\,\lambda$. (d) Variation of the Stokes parameter $V$ in the direction $(\hat{x}-\hat{y})/\sqrt 2$ for different angles of tilt $\theta_i$ and a shift of $x_0=y_0=17.67\,\lambda$. Stokes parameters are plotted in arbitrary units.  The CG of the intensity pattern for LCP  and RCP components are superposed using  white and black dots. The two points coincide in Fig. \ref{fig:V_w_w_out_shift}a.} \label{fig:V_w_w_out_shift}
	\end{figure}
	\subsection{\label{sec:level3.2} Numerical results studying the effects of tilt and shift}
	In the previous section, we saw that a shift between the centers of the beam waist and the aperture with or without tilt results in splitting of the centers of left and right circularly polarized light {associated with opposite topological charges,} which is a '{fingerprint}' of {OOI} of light. We now present the results on the effect of the tilt of the aperture on the {OOI} for a fixed shift.
	\par
\begin{figure}[t]
		\subfloat[\label{fig:V_I_w_w_out_shift:a}]{
			\includegraphics[width=0.48\columnwidth]{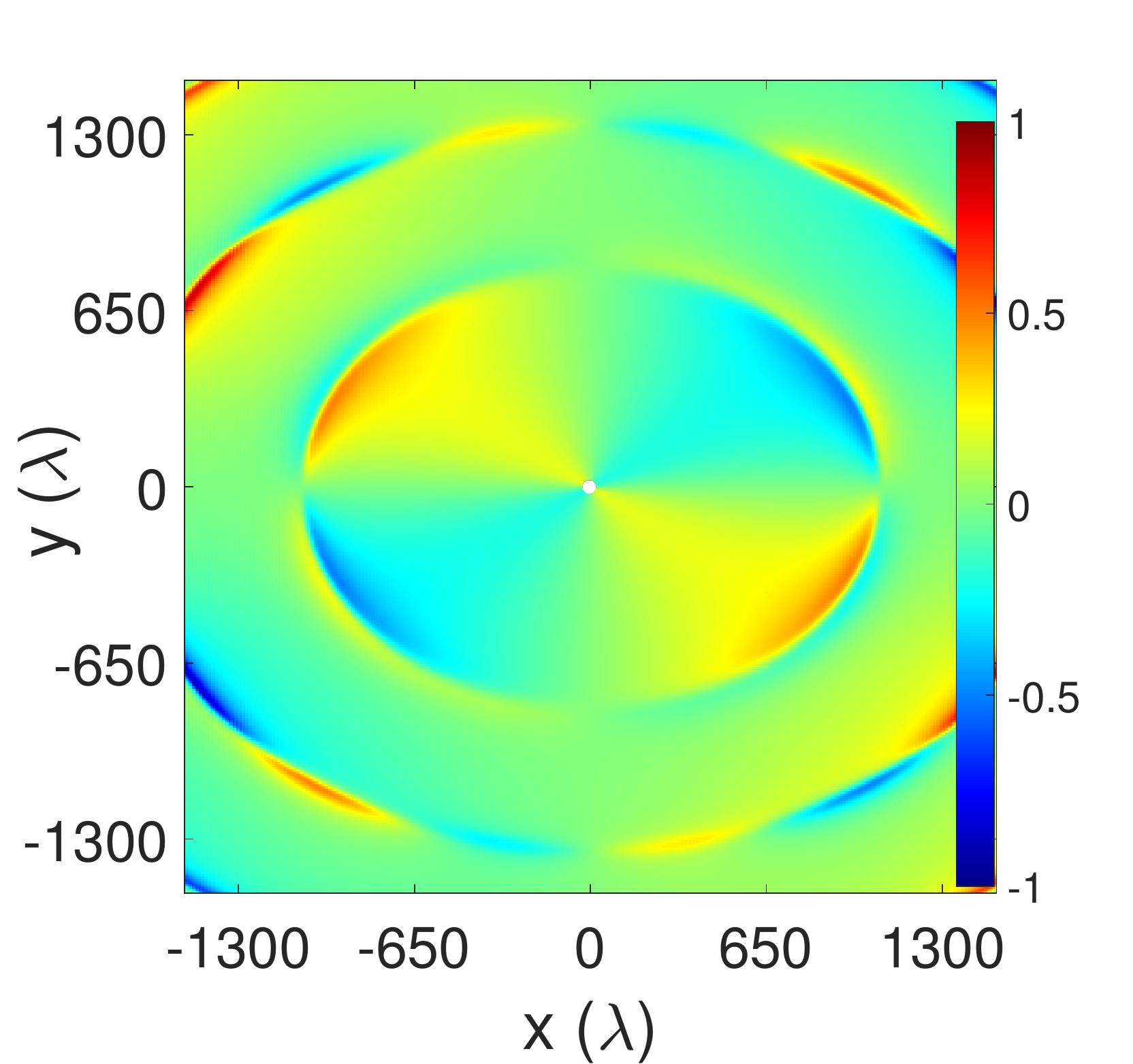}
		}\hspace*{\fill}%
		\subfloat[\label{fig:V_I_w_w_out_shift:b}]{
			\includegraphics[width=0.48\columnwidth]{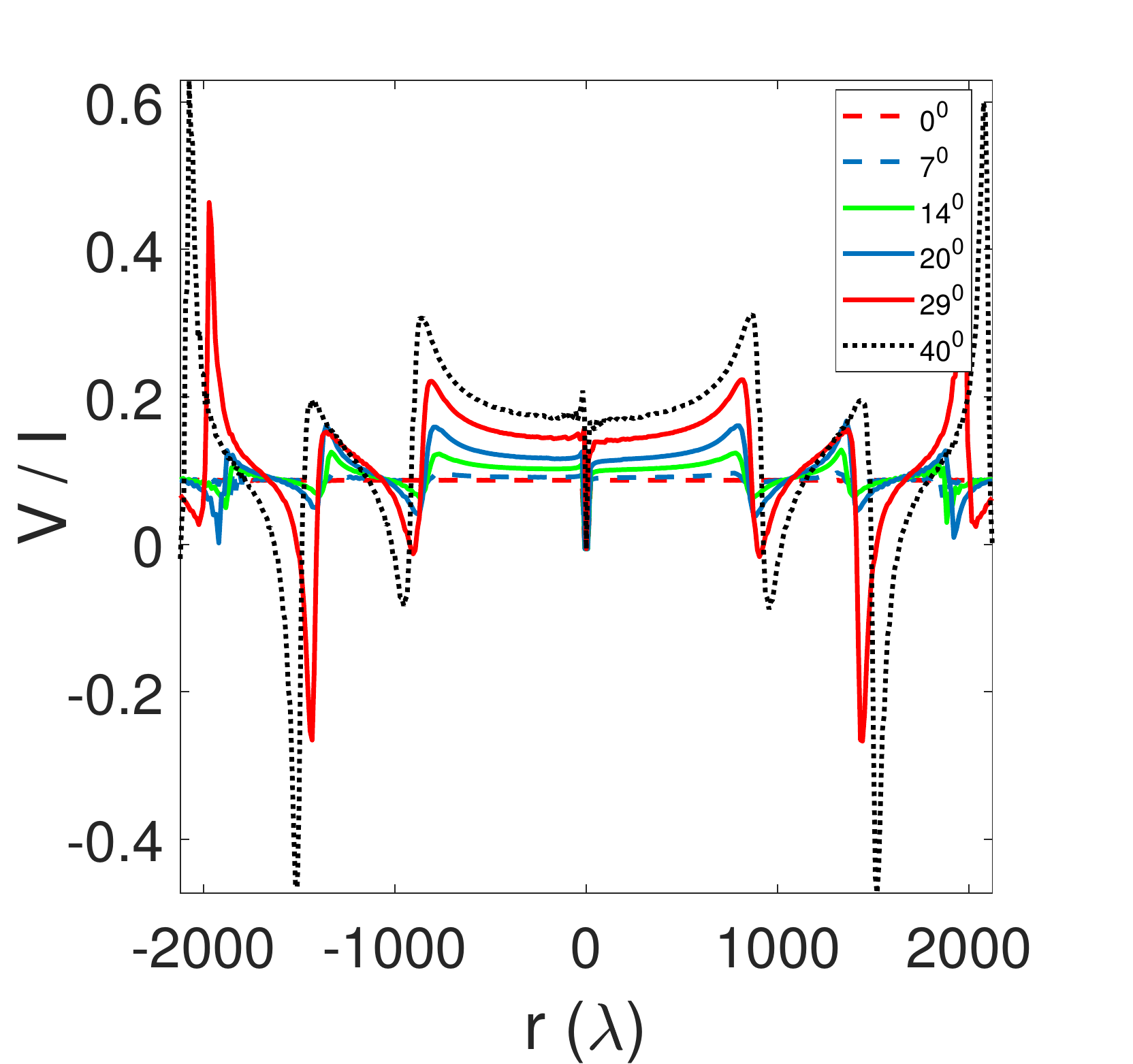}
		}		
	
		\medskip
				
		\subfloat[\label{fig:V_I_w_w_out_shift:c}]{
			\includegraphics[width=0.48\columnwidth]{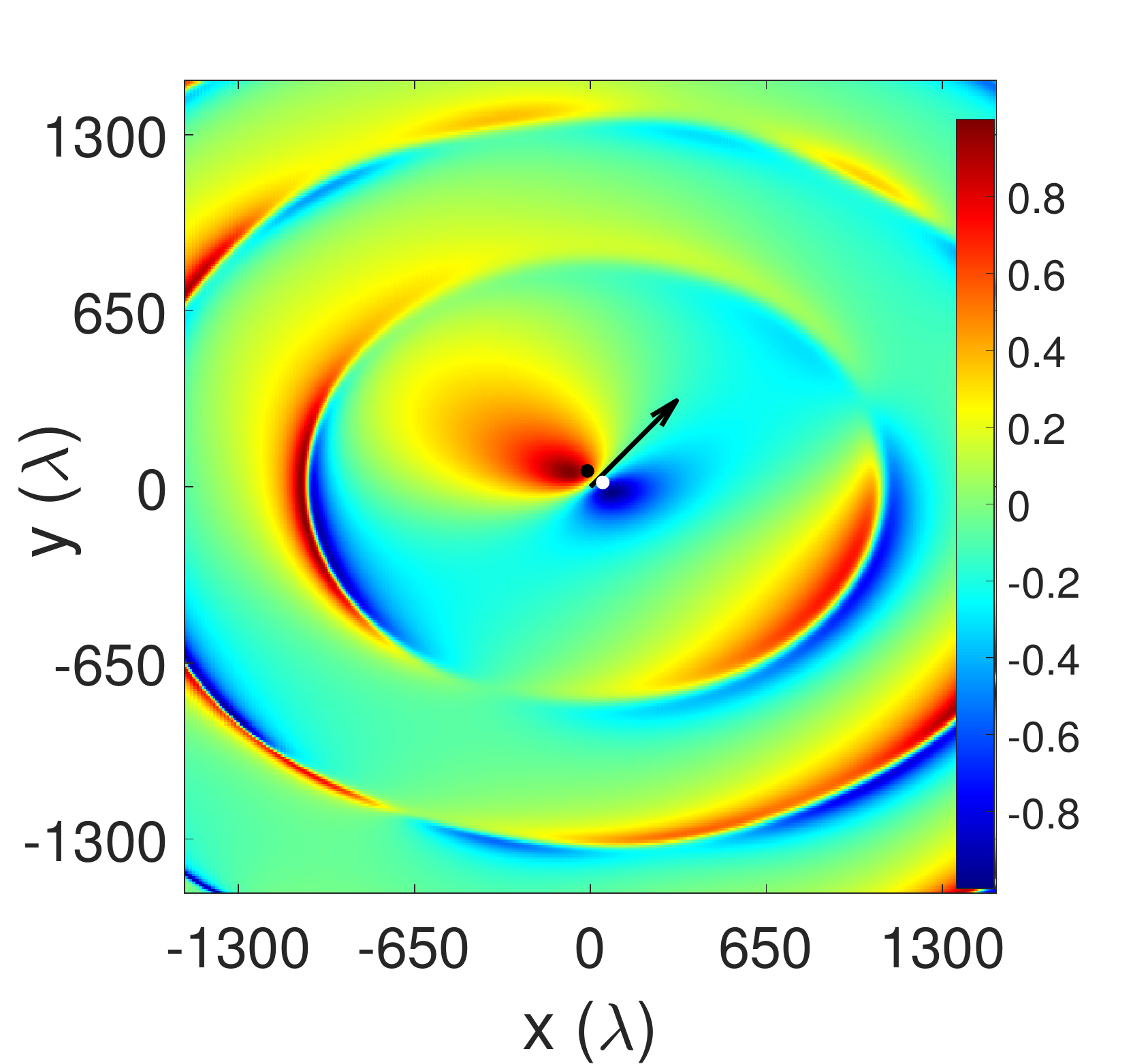}
		}\hspace*{\fill}
		\subfloat[\label{fig:V_I_w_w_out_shift:d}]{%
			\includegraphics[width=0.48\columnwidth]{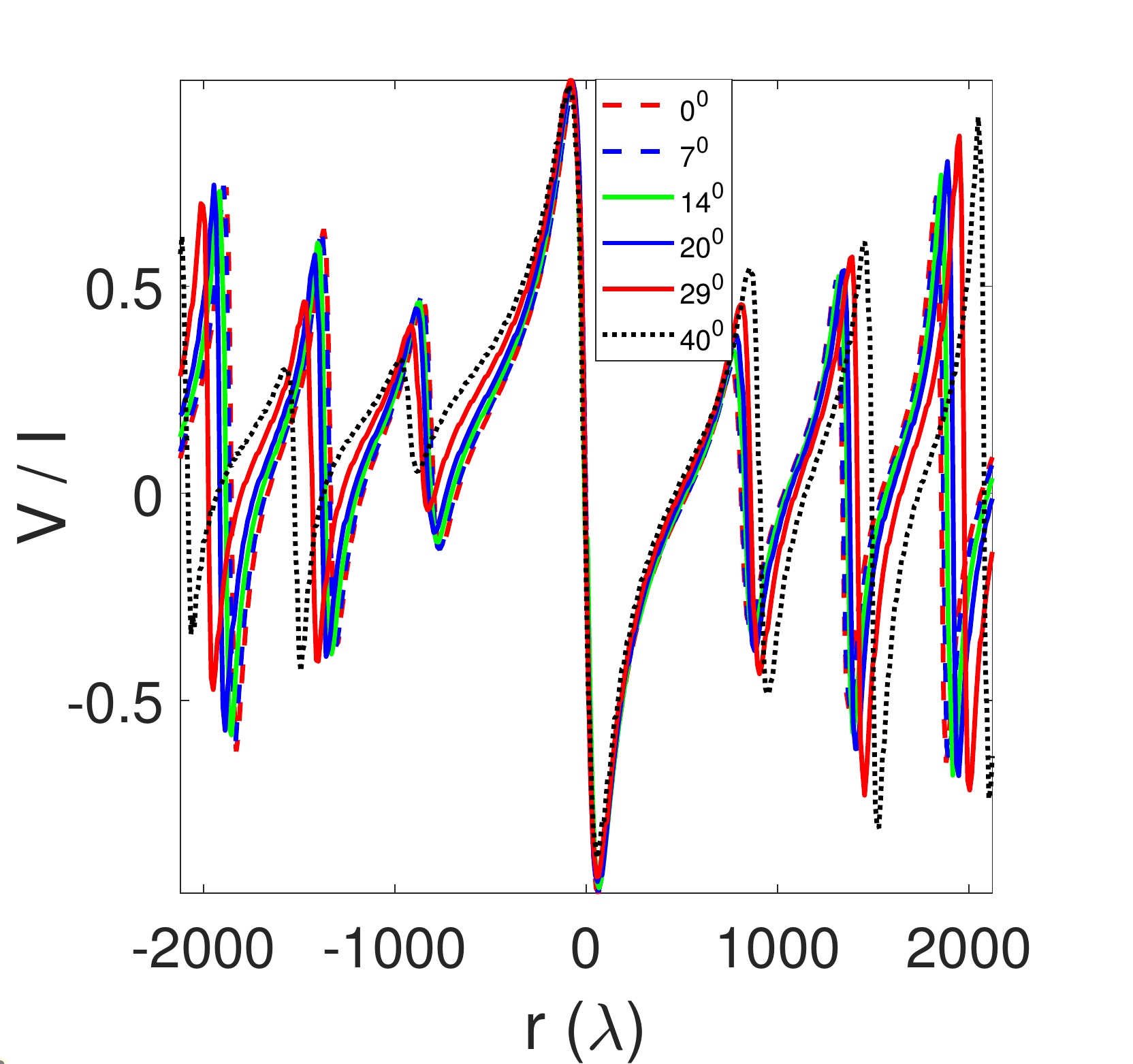}%
		}
		\caption{(a) Numerically calculated degree of circular polarization $V/I$ for a perfectly aligned aperture tilted at an angle $\theta_i=40^0$. (b) Variation of the degree of circular polarization $V/I$ in the direction $(\hat{x}-\hat{y})/\sqrt 2$ for different angles of tilt $\theta_i$ for a perfectly aligned aperture. (c) Numerically calculated degree of circular polarization $V/I$ for an off-axis aperture tilted at an angle $\theta_i=40^0$ and a shift of $x_0=y_0=17.67\,\lambda$. (d) Variation of degree of circular polarization $V/I$ in the direction $(\hat{x}-\hat{y})/\sqrt 2$ for different angles of tilt $\theta_i$ and a shift of $x_0=y_0=17.67\,\lambda$.  The CG of the intensity pattern for LCP  and RCP components are superposed using  white and black dots. The two points coincide in Fig. \ref{fig:V_I_w_w_out_shift}a.} \label{fig:V_I_w_w_out_shift}
	\end{figure}

	We first look at a scenario where the two centers are perfectly aligned, with different tilts for the aperture. {Note that the pattern shown in Fig. \ref{fig:V_w_w_out_shift:a} is affected by the small amount of eccentricity of the bare beam, thus deviating from a perfectly radial vector beam. The variation of the Stokes parameter (or the SAM) in the direction $(\hat x'-\hat y')/\sqrt 2$ for different angles of tilt is shown in Fig.\ref{fig:V_w_w_out_shift:b}.} As can be seen, the tilting increases the amount of SAM, though by a very small amount till about $29^0$. Further tilting reduces the amount of SAM, as depicted by the $\theta_i=40^0$ (Black dotted) curve. {We have superposed the CG of the two helicity components in Figs.\ref{fig:V_w_w_out_shift:a} and \ref{fig:V_w_w_out_shift:c}. It is clear from Fig. \ref{fig:V_w_w_out_shift:a} that a tilt of the aperture in the absence of any shift from the axis clearly does not create a significant change in position of the CG of the intensity for each helicities. When we include a small shift between the two centers, the separation between the CG's is more prominent at $\theta_i=0^0$. In contrast to the previous case, tilting the slit decreases the amount of SAM, as can be seen in Fig. \ref{fig:V_w_w_out_shift:d}.} We also calculated the corresponding degree of circular 	polarization which confirms the same results (see Fig. \ref{fig:V_I_w_w_out_shift}).
		\par
\begin{figure}[t]
		\includegraphics[width=0.95\columnwidth]{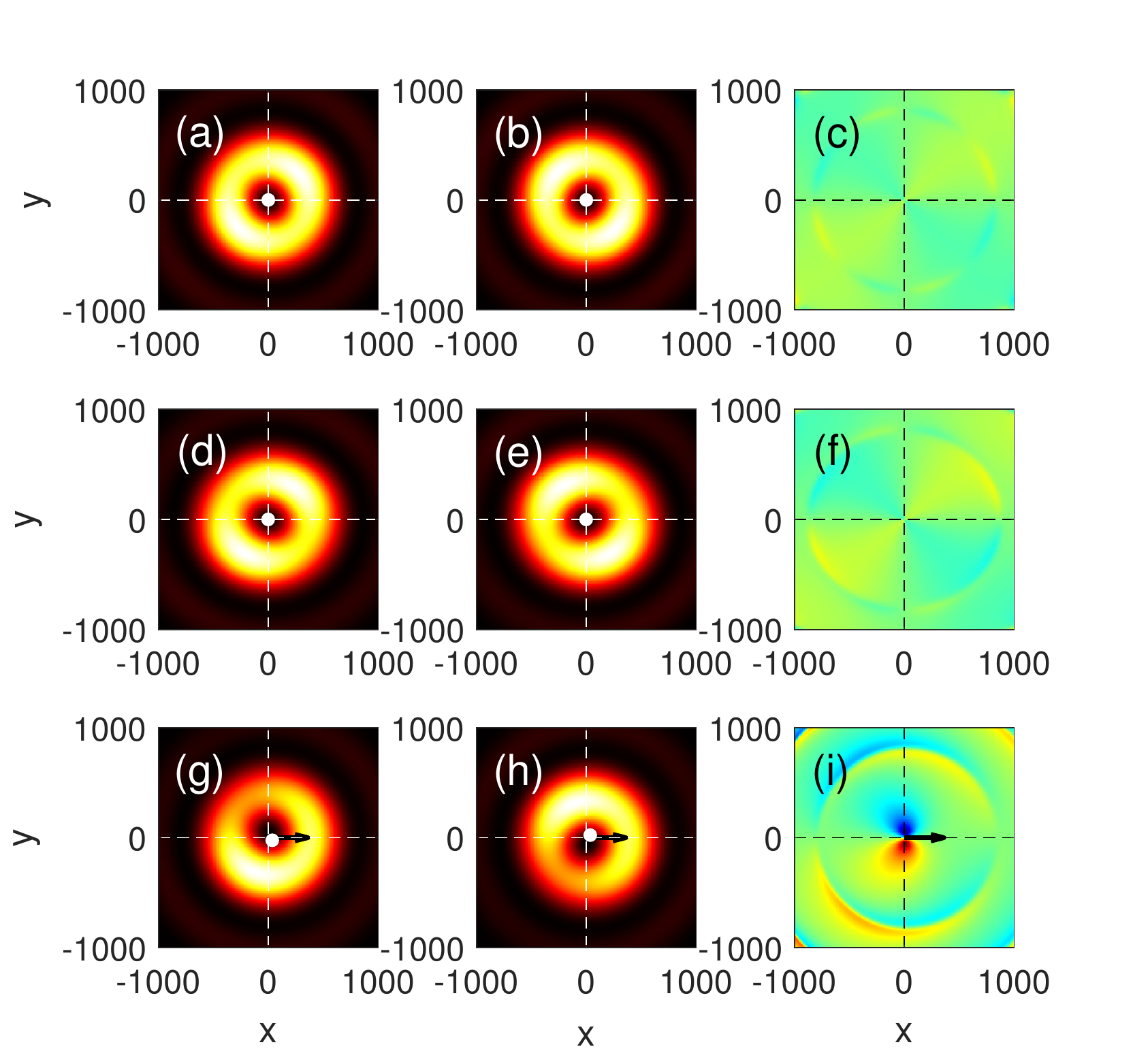}
		\caption{(a), (b) The intensities of the field profile obtained by projecting the diffraction pattern of a beam given by eq. \ref{eq_VB}, into the left and right circular basis. (c) The corresponding Stokes pattern. The shift and tilt of the aperture is given by $x_0 = 0\lambda$, $y_0 = 0\lambda$, and $\theta_i = 0$. (d), (e), (f) Same as (a), (b), (c) except for $x_0 = 0\lambda$, $y_0 = 0\lambda$, and $\theta_i = 20^0$. (d), (e), (f) Same as (a), (b), (c) except for $x_0 = 20\lambda$, $y_0 = 0\lambda$, and $\theta_i = 0^0$. The CG of the distribution is marked using a white dot. In the  bottom panels, the black arrow represents the direction of the shift of the aperture. The corresponding color bars are the same as in Fig. \ref{fig:0deg:a} and \ref{fig:0deg:c}, for the intensity patterns and the Stokes images, respectively.}
		\label{Fig:9}
	\end{figure}

	{The intensity patterns with superposed CG (white dot) along with the Stokes image for the degree of circular polarization for typical parameters are shown in Fig. \ref{Fig:9}. The top row shows the results for a perfectly aligned system with null shift and tilt, while the left, middle and right columns
	 show, respectively, the LCP, RCP and the degree of circular polarization. The middle row gives the same for null shift with $\theta=20^0$, while the bottom row depicts the same for null tilt but with shift $x_0 = 20\lambda$. It is clear from  a comparison of Figs. \ref{Fig:9}a and \ref{Fig:9}b along with panels \ref{Fig:9}d and \ref{Fig:9}e for null shift that { helicity} does not play any role in the behavior of the CG even for tilted systems, while the OO coupling for finite shift leads to opposite movement of the CG, which is also confirmed by the Stokes image (see Fig. \ref{Fig:9}i).}
	\par 

\begin{figure}[t]
		\includegraphics[width=0.95\columnwidth]{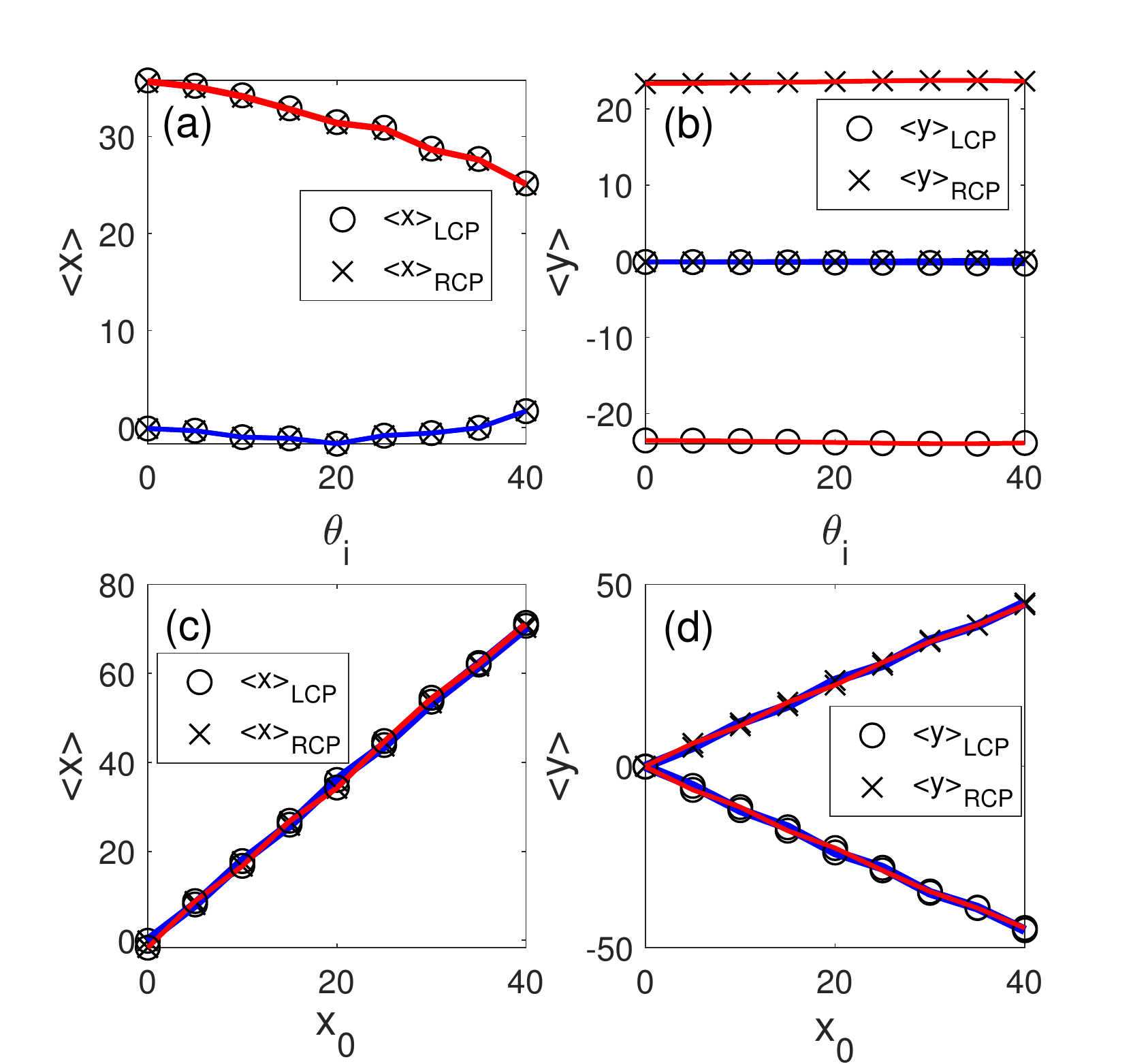}
		\caption{(a),(b) The coordinates of CG $<x>$ and $<y>$ as a function of  the tilt angle $\theta_i$, for $y_0 = x_0=0\lambda$ (blue line) and for $y_0 = 0 \lambda$, $x_0 = 20\lambda$ (Red line). (c), (d)  $<x>$ and $<y>$ as a function of  the displacement from the center $x_0$, for $y_0 =0\lambda$. The tilt angle has been taken to be $\theta_i= 0^0$ (blue line) and  $\theta_i = 20^0$(Red line). Circles (Crosses) represent the CGs of intensity field after projecting onto the LCP (RCP) basis.}
		\label{Fig:10}
	\end{figure}

	{A more detailed picture is presented in Fig. \ref{Fig:10}, where the top (bottom) row shows the variation of position of the CG as  functions of the tilt angle (shift) keeping the shift (tilt angle) fixed at two values.	The left (right) panel represents the $x$ ($y$) component of the CG. Circles (crosses) represent the CG of the intensity pattern projected into the LCP (RCP) basis. The $\theta$ dependence on the difference between the CG of the two polarizations become nearly indistinguishable in absence of any shift (Crosses and circles nearly overlap in the blue curves in Fig. \ref{Fig:10}a,b). Note that if the incident vector beam is ideally radial ($\theta = \delta =0$), then the two components are indistinguishable (not shown). The $y$ component of the CG shows almost no dependence on the tilt angle for either polarizations for both the values of shift (Fig. \ref{Fig:10}b). This is not surprising, since we have chosen the $y$ axis as our axis of rotation. Its small dependence on $\theta_i$ for null shift vanishes completely for a perfect radial beam, as expected.  Interestingly, the  behavior of the corresponding $x$ component shows a minimum for both polarizations, as can be seen from the blue curve in Fig. \ref{Fig:10}a. The effect of the tilt becomes much more pronounced in presence of an off-axis aperture, as can be seen from the  red curve in Fig. \ref{Fig:10}(a), where $<x>$ has shifted by almost $10\lambda$ from its initial value upon tilting the aperture by $40^0$. Thus, for an an off-axis aperture, the notable tilt angle dependence of the CG can become relevant in typical realistic systems. The picture is somewhat different in case of CG as  functions of shift for a fixed angle $\theta_i$ (Fig. \ref{Fig:10}c and \ref{Fig:10}d). The x-component of the CG is positive with almost identical linear dependence  on the aperture--beam-center offset $x_0$ for both the helicities as well as for both tilt angles (Fig. \ref{Fig:10}c). In contrast, the $y$ component is helicity dependent, and goes in opposite directions with same results for the two tilt values (Fig. \ref{Fig:10}d). Recall that the helicity dependence of the CG of the diffraction pattern of a vector beam is actually an effect of the different $l$ values associated with each helicity.}
		
		
		
	\section{\label{sec:level4}Conclusions}
	In this article, we have studied the diffraction of an off-axis vector beam from a tilted aperture. {The  broken symmetry is shown to result in mixing of the intrinsic and extrinsic orbital angular momentum of light, leading to a split between the LCP and RCP components of the incident beam. Note that these polarization components are associated with  opposite topological charges.} We report the increase in this splitting with increasing separation between the beam and aperture centers. {Our numerical results have been validated using both experimental techniques, as well as using brute force integration. We also shed light on the connection between the spin angular momentum density and the Stokes parameter $V$. We have presented a thorough analysis of the individual and combined effects of shift and tilt on the diffraction pattern, on the corresponding Stokes images and also on the center of gravity of the polarization components.}



	\begin{acknowledgments}
		We wish to acknowledge the support of the Indian Institute of Science Education and Research (IISER) Kolkata, an autonomous institute under the Ministry of Human Resource Development (MHRD), Govt. of India. We would like to acknowledge Science and Engineering  Research Board (SERB), Govt. of India for funding. We also acknowledge financial support from the Council of Scientific and Industrial Research (CSIR), Government of India. 
		\par
		{GR and ABS  have contributed equally to this work.}
	\end{acknowledgments}
	\bibliography{Diffraction_of_an_Off_axis_Vector_Beam_by_a_Tilted_Aperture}
\end{document}